\documentclass[%
 aip,
 amsmath,amssymb,
 reprint,%
]{revtex4-1}

\usepackage{tabularx}
\usepackage{amsmath}
\usepackage{graphicx}% Include figure files
\usepackage{dcolumn}% Align table columns on decimal point
\usepackage{bm}% bold math
\usepackage[utf8]{inputenc}
\usepackage{caption, subcaption}
\usepackage[T1]{fontenc}
\usepackage{mathptmx}
\usepackage{etoolbox}
\usepackage{array, booktabs}
\usepackage{makecell} 

\makeatletter
\def\@email#1#2{%
 \endgroup
 \patchcmd{\titleblock@produce}
  {\frontmatter@RRAPformat}
  {\frontmatter@RRAPformat{\produce@RRAP{*#1\href{mailto:#2}{#2}}}\frontmatter@RRAPformat}
  {}{}
}%
\makeatother
\begin{document}

\preprint{AIP/123-QED}

\title[Influence of electrical properties on thermal boundary conductance at metal/semiconductor interface]{Influence of electrical properties on thermal boundary conductance at metal/semiconductor interface}
% Force line breaks with \\

\author{Q. Pompidou}
 \affiliation{%
ITheMM, Université de Reims Champagne-Ardennes URCA, Moulin de la Housse, BP 1039, 51100 Reims, France
}%

\author{C. Acosta}
	\affiliation{%
INSA de Lyon, CETHIL, UMR5008, 69621 Villeurbanne, France
}%

\author{M. Brouillard}
\affiliation{
Univ. Lille, CNRS, Univ. Polytechnique Hauts-de-France, Junia, UMR 8520 - IEMN - Institut d'Electronique de Microélectronique et de Nanotechnologie, F-59000 Lille, France
}%

\author{N. Bercu}
\affiliation{
L2n, UMR CNRS 7076, 12 rue Marie Curie, Université de Technologie de Troyes, 10004 Troyes, France
}%

\author{L. Giraudet}
\affiliation{
L2n, UMR CNRS 7076, 12 rue Marie Curie, Université de Technologie de Troyes, 10004 Troyes, France
}%  

\author{R. Sheikh}
	\affiliation{%
Department of Mechanical Engineering and Materials Science, University of Pittsburgh, Pittsburgh, Pennsylvania, USA
}%
\author{C. Adessi}
	\affiliation{%
Institut Lumière Matière, UMR5306, Université Claude Bernard Lyon 1-CNRS, Université de Lyon, Villeurbanne, 69622, France
}%
\author{S. Mérabia}
	\affiliation{%
Institut Lumière Matière, UMR5306, Université Claude Bernard Lyon 1-CNRS, Université de Lyon, Villeurbanne, 69622, France
}%
 \author{S. Gomès}
	\affiliation{%
INSA de Lyon, CETHIL, UMR5008, 69621 Villeurbanne, France
}%

\author{P.-O. Chapuis}
	\affiliation{%
INSA de Lyon, CETHIL, UMR5008, 69621 Villeurbanne, France
}%

\author{J.-F. Robillard}
\affiliation{
Univ. Lille, CNRS, Univ. Polytechnique Hauts-de-France, Junia, UMR 8520 - IEMN - Institut d'Electronique de Microélectronique et de Nanotechnologie, F-59000 Lille, France
}%

\author{M. Chirtoc}
 \affiliation{%
ITheMM, Université de Reims Champagne-Ardennes URCA, Moulin de la Housse, BP 1039, 51100 Reims, France
}%
\author{N. Horny}%
  \affiliation{%
ITheMM, Université de Reims Champagne-Ardennes URCA, Moulin de la Housse, BP 1039, 51100 Reims, France
}%
 \email{nicolas.horny@univ-reims.fr}

\date{\today}% It is always \today, today,

\begin{abstract}
    Recent experimental investigations have demonstrated that doping a semiconductor is a route to increase the thermal boundary conductance at metal/semiconductor interfaces. In this work, the influence of the electrical properties on heat transfer across metal/doped semiconductor junctions is investigated. Specifically, thermal boundary conductance at the interfaces between $p$ and $n$ doped silicon and titanium is measured by employing frequency domain photothermal radiometry under varying external conditions. The influence of the doping level of the semiconductor, the barrier height and the space charge area is analyzed. In particular, a 40 percent increase of the interface thermal conductance with the application of a current at $n$-doped silicon/titanium interfaces is reported. The enhancement of the thermal boundary conductance is explained by the shrinking of the surface charga area induced by the electric current. This study opens the way to modulating interfacial heat transfer at metal/semiconductor interfaces through fine tuning of electrical effects.
\end{abstract}

\maketitle
\section{Introduction, overview and challenges}\label{Introduction}

    With the miniaturization of electronic devices and their growing complexity, thermal management has emerged as a major issue for new technologies. Thermal issues refer to the challenges and consequences associated with heat generation and dissipation. Given the increase in power densities, specifically in nanoscale electronics, thermal management is essential to prevent overheating, which can lead to breakdowns, reduced performances, and even irreversible failures. One of the main challenges concerns the interfaces between component materials, which can have very low thermal boundary conductance (TBC), thus impeding heat flow. This phenomenon can be observed in particular due to the considerable increase in interface density caused by the growing number of nanometric elements in modern components. While nanoscale engineered interfaces are developed to solve this thermal problem, heat transfer at metal-semiconductor (M-SC) interfaces remains poorly understood. This is the main focus of the experimental work that specifically analyzes the electronic contribution to heat transport across these interfaces when bias voltage is applied to the M-SC junction as in operating components. 
The TBC $G$ is a physical quantity defined as the ratio between the applied heat flux and the temperature drop at the junction. The general macroscopic expression governing this phenomenon is defined as follows: 

\begin{equation}
\label{TBC_G}
    \phi = G \Delta T
\end{equation}

with $G$ the TBC in $\mathrm{W \cdot m^{-2} \cdot K^{-1}}$, $\phi$ the heat flux in $\mathrm{W\cdot m^{-2}}$ and $\mathrm{\Delta} T$ the temperature drop at the interface in K.

This TBC depends on the transmission of heat carriers across the interface and therefore depends on the thermal properties of both materials. If the thermal conduction is mainly done by phonons in both materials, the interface’s TBC refer to phonon transport. However, when one of the materials is a metal, the electrons could take an important part of the thermal transfer, and the energy transfer at the interface is not as intuitive. 
	
\begin{figure}[h!]
	\centering
	\includegraphics[width=0.40\textwidth]{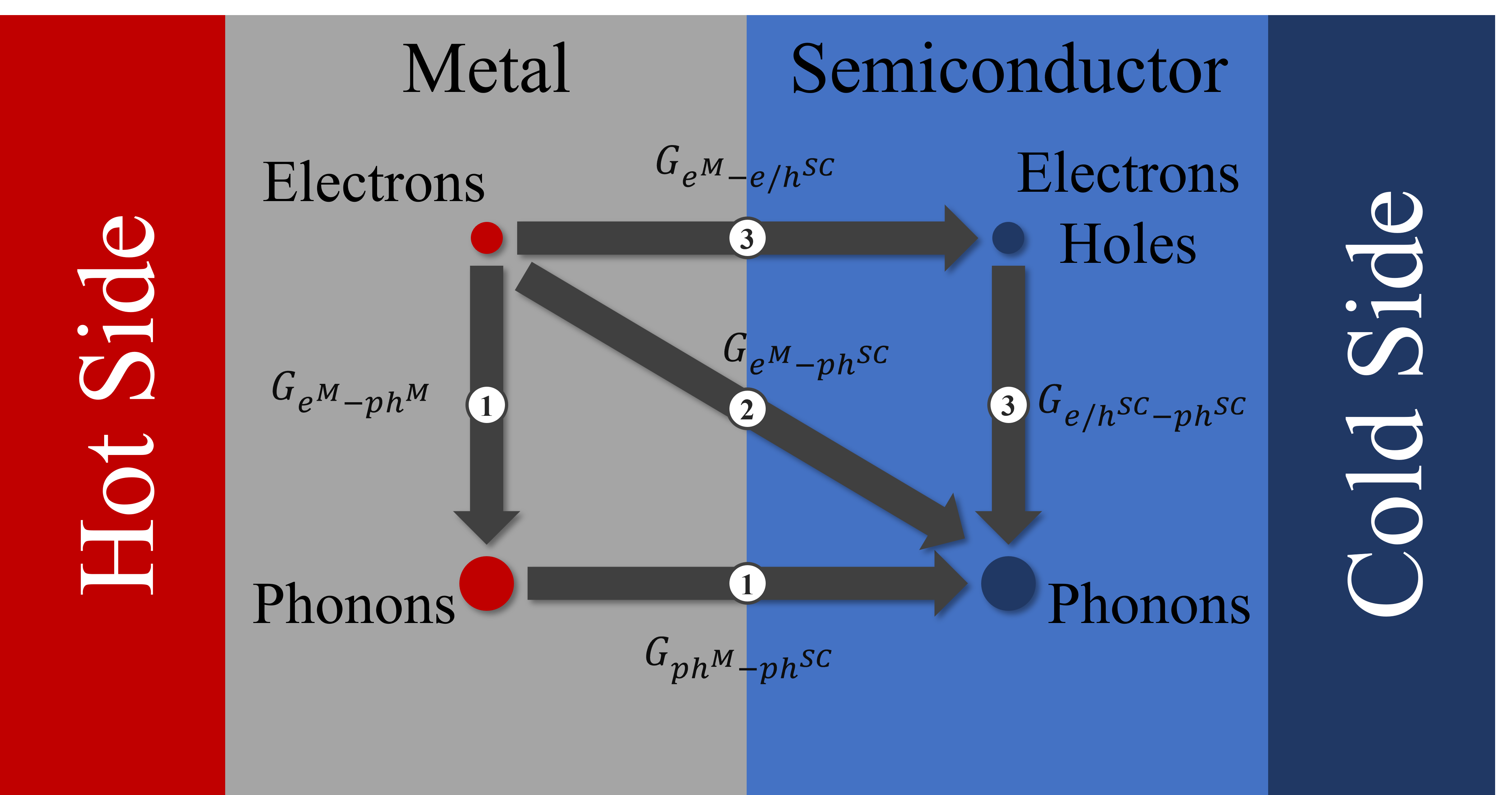}
	\caption{Schematic representation of the three possible channels of transmission for the heat flux through a metal/semiconductor interface.}\label{fig:pathways}
\end{figure}

A study by Majumdar and Reddy \cite{majumdar_role_2004} concluded that electron-phonon coupling in the metal and then phonon transmission through the interface (channel (1) in Figure \ref{fig:pathways}) was not sufficient to explain the phenomenon. Thus, electronic transport (channel (2)) must be taken into account as a possible pathway through a junction of conductive/insulator material. Later, Lombard \textit{et al.} \citep{lombard_influence_2014} used a combination of a two-temperature numerical model and analytical works to confirm this hypothesis. They assumed that thermal transport coming from direct coupling between the electrons of the metal and the phonons of the nonmetal was not negligible. In case of semiconductors, a third channel of conduction through the interface could be considered because of the presence of electrical charges in doped systems. Consequently, thermal conduction at metal/semiconductor junctions is a combination of contributions from different pathways as illustrated in Figure \ref{fig:pathways} and can be summarized as follows:
	
i) inside the metal, the electrons give their energy to crystal phonons, then conduction across the interface is mediated by metal phonons-SC phonons coupling ($G_{ph^{M}-ph^{SC}}$); ii) direct metal electrons to SC phonons coupling via $G_{e^{M}-ph^{SC}}$; iii) metal electrons give their energy to charges present in SC, near the interface, and these charges relax within the SC crystal ($G_{e^{M}-e/h^{SC}}$ then $G_{e^{SC}-ph^{SC}}$). The overall thermal transport is described by the TBC $G$ defined in eq. \eqref{TBC_G}. \newline

A recent first-principle study unveiled the impact of electronic effects on interfacial heat transfer at metal/semiconductor interfaces \citep{feliciano2025firstprinciples}. In this study, it was shown that the TBC is generally dominated by phonon-phonon scattering processes at the interface, and that electrons contribute to interfacial heat transfer when the metal’s Debye frequency approaches that of silicon. This contribution involves an electron-phonon coupling localized in the very vicinity of the interface (first atomic layer).

The value of TBC values between metals and semiconductors are well documented in the literature \cite{stevens_measurement_2005, giri_mechanisms_2015, hamaoui_electronic_2018, horny_kapitza_2016, blank_influence_2019, blank_towards_2019, giri_review_2020} and range from $50$ to $360\,\text{ MW}\cdot\text{m}^{-2}\cdot \text{K}^{-1}$ depending on the metal/SC couple. These values are also highly dependent on the deposition processes, as this affects the interfacial strengths. In this work, we present a comprehensive methodology for further characterizing experimentally heat transfer at M/SC interfaces and to analyze the influence of electrical properties on TBC. The relevant electric quantities that influence the flow of electrons at the interface are: (1) the Schottky barrier height (SBH), which is the energy barrier at the interface for electrical carriers; (2) the doping level of the semiconductor $n_D$, that we tune in this study from intrinsic to highly doped SC, and (3) the electrical potential $V$ applied to the junction, also named operando bias in this article. Thus, the TBCs are measured and analyzed with respect to these three main electrical parameters. It should be noted that if surface polaritons, which exist at metal-semiconductor interface, have any effect on the transport perpendicular to the interface, this effect is embedded in the thermal boundary conductance measured experimentally.
Titanium or platinium /doped silicon contacts are studied as M/SC contacts. The objective is to understand the electronic contribution to thermal transport across these junctions when they are polarized. The samples specifically designed for the study are first described. We then present electrical characterizations before to analyze the M/doped Si contacts TBCs, measured by means of frequency domain photothermal radiometry (FD-PTR) technique.

\section{Sample design}\label{Sample_design_sec}

\begin{figure}[h!]
	\includegraphics[width=0.4\textwidth]{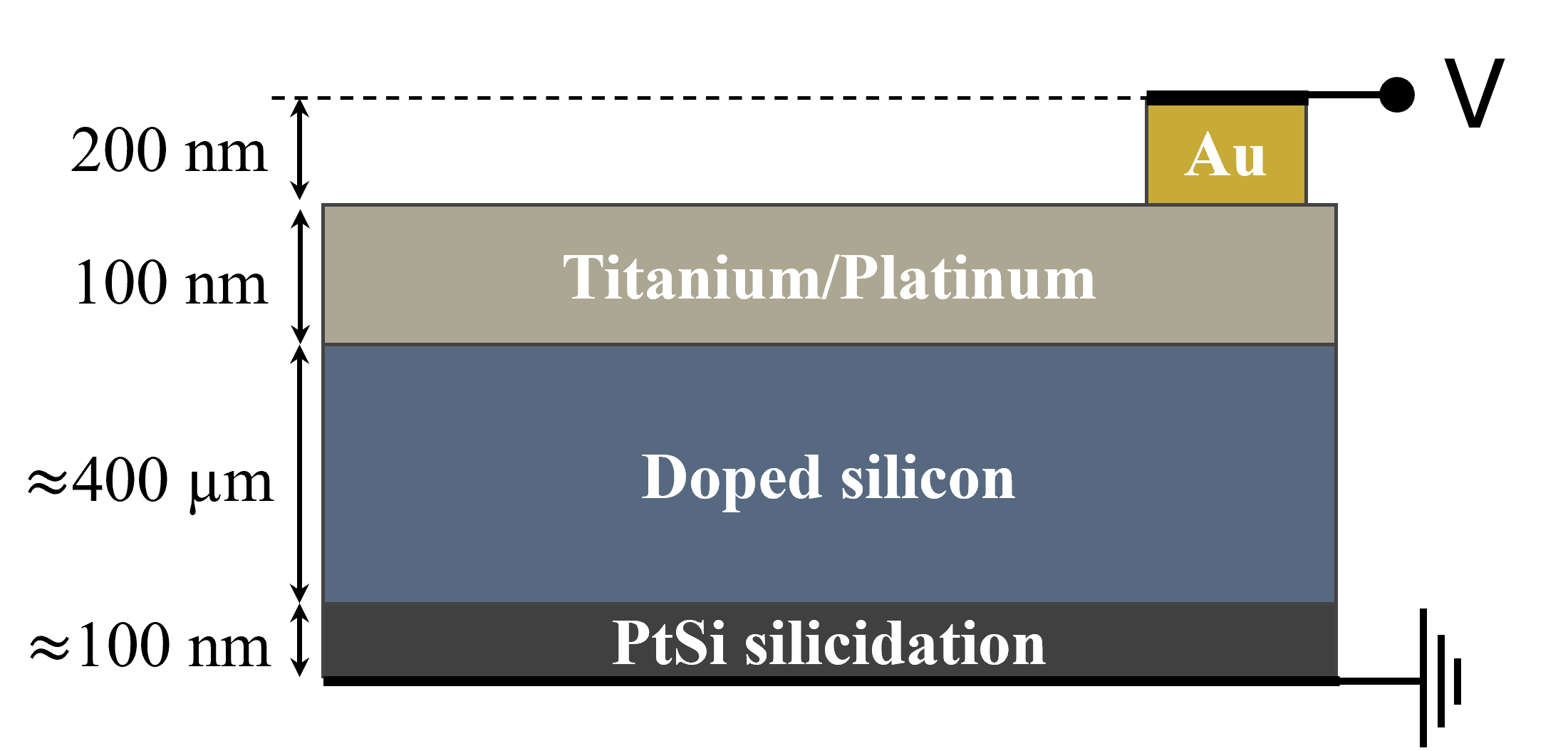}
	\caption{\label{fig:Sample}  Schematic representation of samples showing the convention of polarization adopted throughout this work.} 
\end{figure}

The batch of studied samples consists of silicon substrates with different levels of doping, covered with a titanium or platinum layer having a disk shape. Then, ring-shaped gold electrodes were deposited on these metal disks to polarize the M/SC junction during TBC measurements (see Figure \ref{fig:Sample}). All depositions were performed by electron-beam physical vapor deposition (EB-PVD) to provide regular surfaces.
To ensure consistent surface conditions across all samples, the preparation process was standardized. The silicon substrates were first cleaned in a sulfuric acid ($H_2 SO_4$) bath for 10 minutes. This was followed by a 2-minute rinse in deionized water, after which the substrates were dried with nitrogen ($N_2$). Next, the substrates were immersed in a 1 $\%$ hydrofluoric acid (HF) bath for 10 minutes. Finally, they underwent another 2-minute rinse in deionized water and were dried again with nitrogen ($N_2$). To further prepare the surface, the substrates were subjected to argon etching for 2 minutes at 200 eV before metal deposition. This sequence ensured uniformity in the surface preparation of all the samples and the absence of a layer of silicon oxide at the interface.
Seven samples were produced with 100 nm titanium on (100) silicon substrates with different doping levels (Table \ref{data_table}): low, intermediate and strong levels. In addition, five other samples were made with platinum instead of titanium. On these last five samples, only the variation of TBC was studied according to doping level. The thickness of 100 nm was choosen to obtain the best compromise between sensitivity to the TBC and signal to noise ratio during IR-PTR measurements.
The bias/electrical current application was achieved by means of a Power Supply Unit (PSU) via wire-bonding connected on gold electrodes, and monitored by means of a multimeter for accurate control of the current/potential.

\begin{table}[h]
\caption{Electrical characteristics of the different silicon substrates. The dopant density has been provided by the manufacturer of the silicon wafers. NID is for Not Intentionally Doped.}\label{data_table}
\centering
\small

  \begin{tabular}{|c|c|c|}	
    \toprule
     $\,$ Doping type $\,$ &    \makecell{$\,$ $\,$ Dopant density $n_{\text{D}}$ $\,$ $\,$\\ $[\text{cm}^{-3}]$}    &    \makecell{$\,$ $\,$ Electrical resistivity $\rho$ $\,$ $\,$\\ $[\Omega\cdot \text{cm}]$} \\
		\hline
        NID-type		&	$<1.0\cdot 10^{13}$	&	$>200$  \\
		$n$-type			&	$4.5\cdot 10^{14}-9.0\cdot 10^{14}$	&	$5-10$  \\
		$p$-type			&	$1.4\cdot 10^{15}-2.3\cdot 10^{15}$	&	$5-10$  \\
		$n^{+}$-type		&	$3.0\cdot 10^{17}-9.0\cdot 10^{17}$	&	$0.02-0.04$\\
		$p^{+}$-type		&	$4.0\cdot 10^{17}-8.0\cdot 10^{17}$	&	$0.05-0.07$\\
        $n^{++}$-type		&	$>1.9\cdot 10^{21}$	&	$<0.005$\\
       	$p^{++}$-type		&	$>2.1\cdot 10^{21}$	&	$<0.005$\\ 

		\toprule
	\end{tabular}

\end{table}

The convention of polarization (potential applied on the metal film) is represented on Figure \ref{fig:Sample} and fixed throughout this work. The functionalization of the PtSi backside is described in Section \ref{I(V)_sec}. The geometry has been optimized to obtain a homogeneous distribution of the current density at the interface wherever the TBC measurements are performed. These considerations justify the choice of a circular-shaped design for the metal coating.

\begin{figure}
    \centering
    \includegraphics[width=0.45\textwidth]{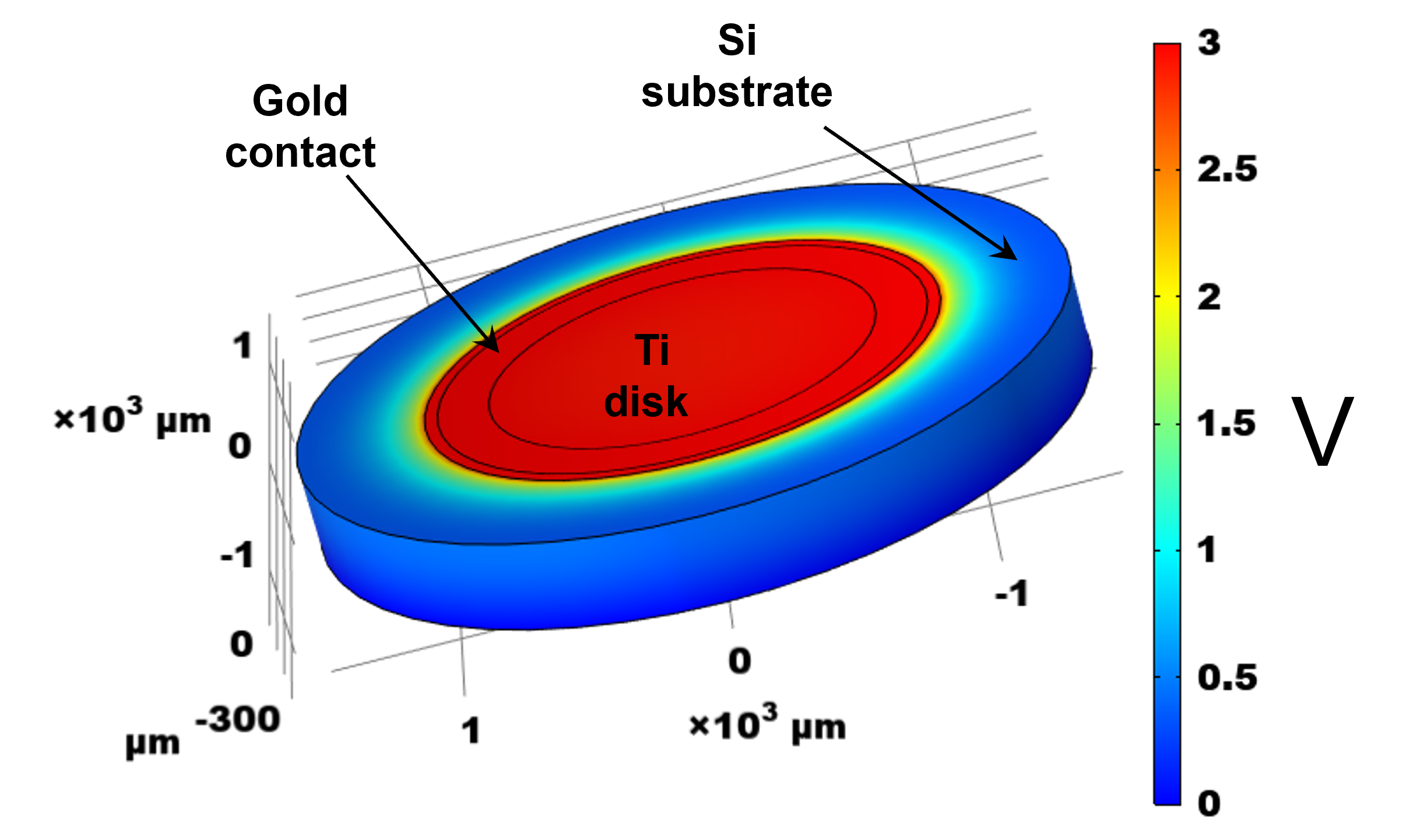}
    \caption{3D simulations of the potential within the sample using the finite element method.}
    \label{fig:COMSOL_3D_Simu}
\end{figure}

In parallel, finite element method (FEM) simulations were performed (COMSOL) to validate the electric behavior of the different devices and to obtain the distribution of the current densities in the depth of the experimental component. One simulation is represented in Figure \ref{fig:COMSOL_3D_Simu}, where the circular shape device is shown with a +3 V potential applied to gold electrodes. The parameters of the simulations are taken according to the manufacturer's silicon substrate datasheets provided in the Table \ref{data_table} for $n$ and $p$-type. The Shottky barrier height was taken as $\phi_B \approx$ 0.5 eV for a Ti/Si junction \citep {dubois_measurement_2004,lee_theoretical_2017}. 

\begin{figure}
    \centering
    \includegraphics[width=0.48\textwidth]{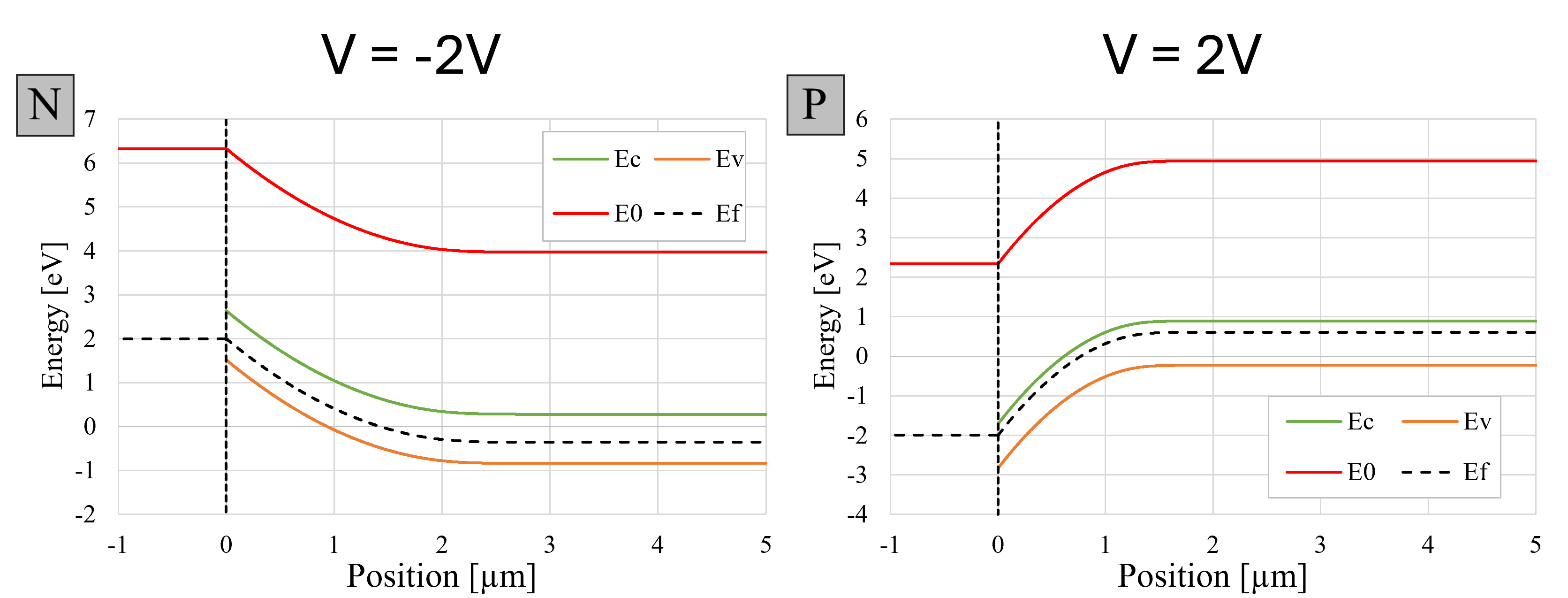}
    \caption{Band diagrams extracted from simulations for $n$- (left) and $p$- (right) doping level substrates, with a 2 V reverse bias applied to the M/SC junction. Solid green and orange curves refer to the conduction and valence bands respectively, dashed lines represent the Fermi level and $E_0$ is the work of extraction. On each figure, the metal is located at the left side.}
    \label{fig:band_diag_comsol}
\end{figure}

From these simulations, the band diagram and the width \textit{W} of the space charge area (or depleted region) were extracted and represented in Figure \ref{fig:band_diag_comsol}.

The results lead to a depleted region $W$ of $1.5$ µm for the $p$-doped silicon and $W= 2.5$ µm for the $n$-doped sample for a $\pm$2 V bias voltage. These values of the spatial extensions are consistent with the $C$($V$) measurements that will be presented in the Section \ref{sec:CV_results}.

\section{Electrical properties of Ti/doped Si interfaces}\label{elec_charac_sec}

In order to investigate correlations between the electrical properties and the thermal properties, the electrical properties (barrier height $\phi_B$, doping level $n_D$ and space charge width $W$) of Ti/doped Si junctions were determined from measurements of current-bias characteristics as a function of temperature $I$($V$, $T$) and capacitance-bias characteristics $C$($V$) of the samples.

    \subsection{Current-Bias $I$($V$) characterization}\label{I(V)_sec}

\begin{figure}
    \centering
    \includegraphics[width=0.40\textwidth]{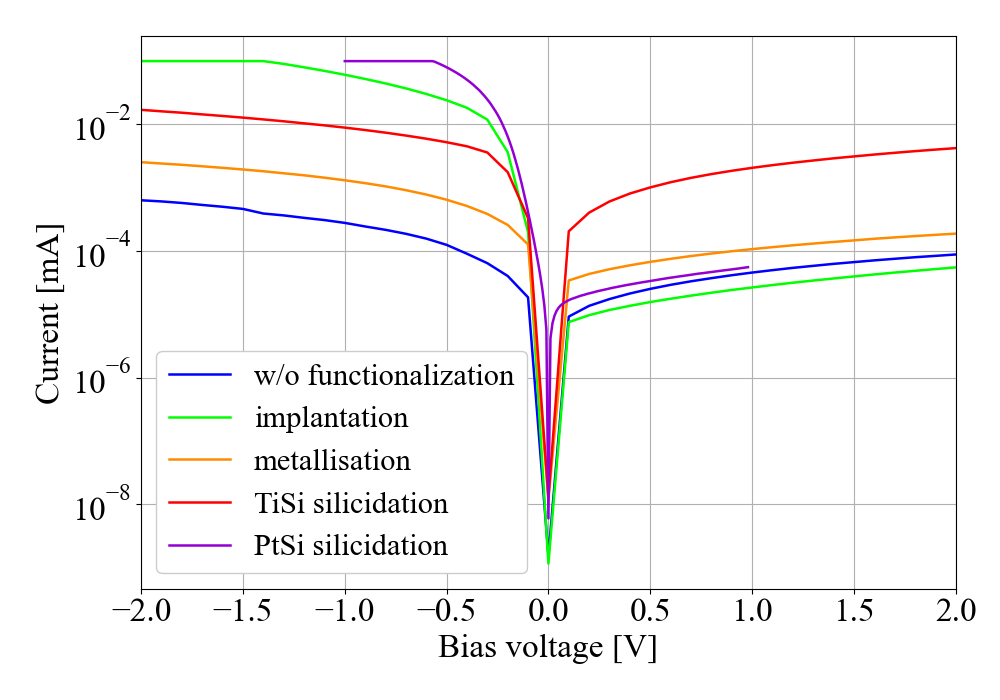}
    \caption{Current-bias characteristics of the titanium/$p$-doped silicon sample for the different methods tested for the rear functionalization of silicon substrates.}
    \label{fig:rear_IV}
\end{figure}

The current-bias characteristics of the Schottky contact were established by polarizing samples by means of a microprobe and collecting the current flowing through each of the devices. 
In order to avoid an influence of the Schottky contact on the rear face of the sample, we tested different passivation processes on this rear face: i) ionic implantation; ii) metallization; iii) Ti silicidation; iv) Pt silicidation. The $I$($V$) characteristics of these different processes are shown in Figure \ref{fig:rear_IV} in the case of $p$-doped silicon substrate.

As can be seen in Figure \ref{fig:rear_IV}, Pt silicidation of the back contact is the best solution to have a Schottky behavior. We have adopted this protocol all the following of our study. However, the Si/PtSi contact, which is an ohmic contact when the silicon is $p$-doped, presents a rectifying behavior when Si is doped with $n$-type inclusions. In this case, the Pt silicidation of the back contact leads to devices composed of two head-to-foot Schottky diodes, and the resulting effect is a Ti/$n$-doped Si with a lower direct current. Nevertheless, as the thermal measurements were performed at high frequencies where the thermal diffusion lengths are shorter than the substrate thickness (around 400 µm), the rear face configuration is not an issue for M/doped Si TBC measurements.

\begin{figure}
    \centering
    \includegraphics[width=0.40\textwidth]{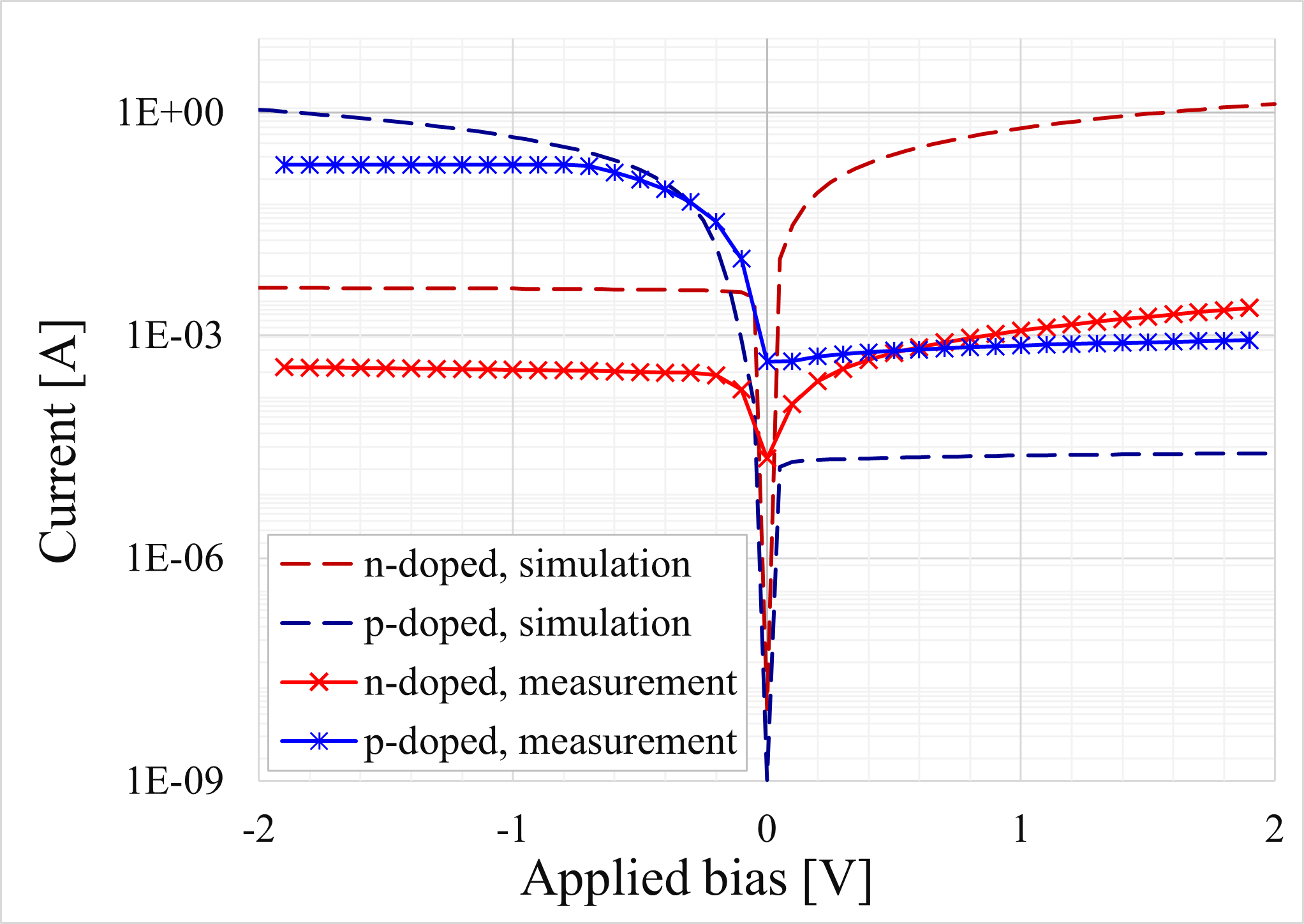}
    \caption{Comparison of the measured and simulated current-bias $I$($V$) characteristic for the titane/$n$- and $p$- doped silicon samples.}
    \label{fig:I(V)_COMSOL_Exp}
\end{figure}

The $I$($V$) measurements performed on $p$- and $n$-doped silicon substrates were compared with the COMSOL Multiphysics simulations and the results are presented in Figure \ref{fig:I(V)_COMSOL_Exp}.

As expected, the behavior between the model and the experiment is comparable for both types of dopant ($n$ and $p$). Under reverse mode, the real residual current is higher mainly because of the leakage current, and as explained above, the Ti/$n$-type silicon shows a direct current below the predicted value because of the rear contact of the device that presents a Schottky behavior.

\begin{figure}[h]
	\centering
	\begin{subfigure}{0.235\textwidth}
	   \includegraphics[width=\textwidth]{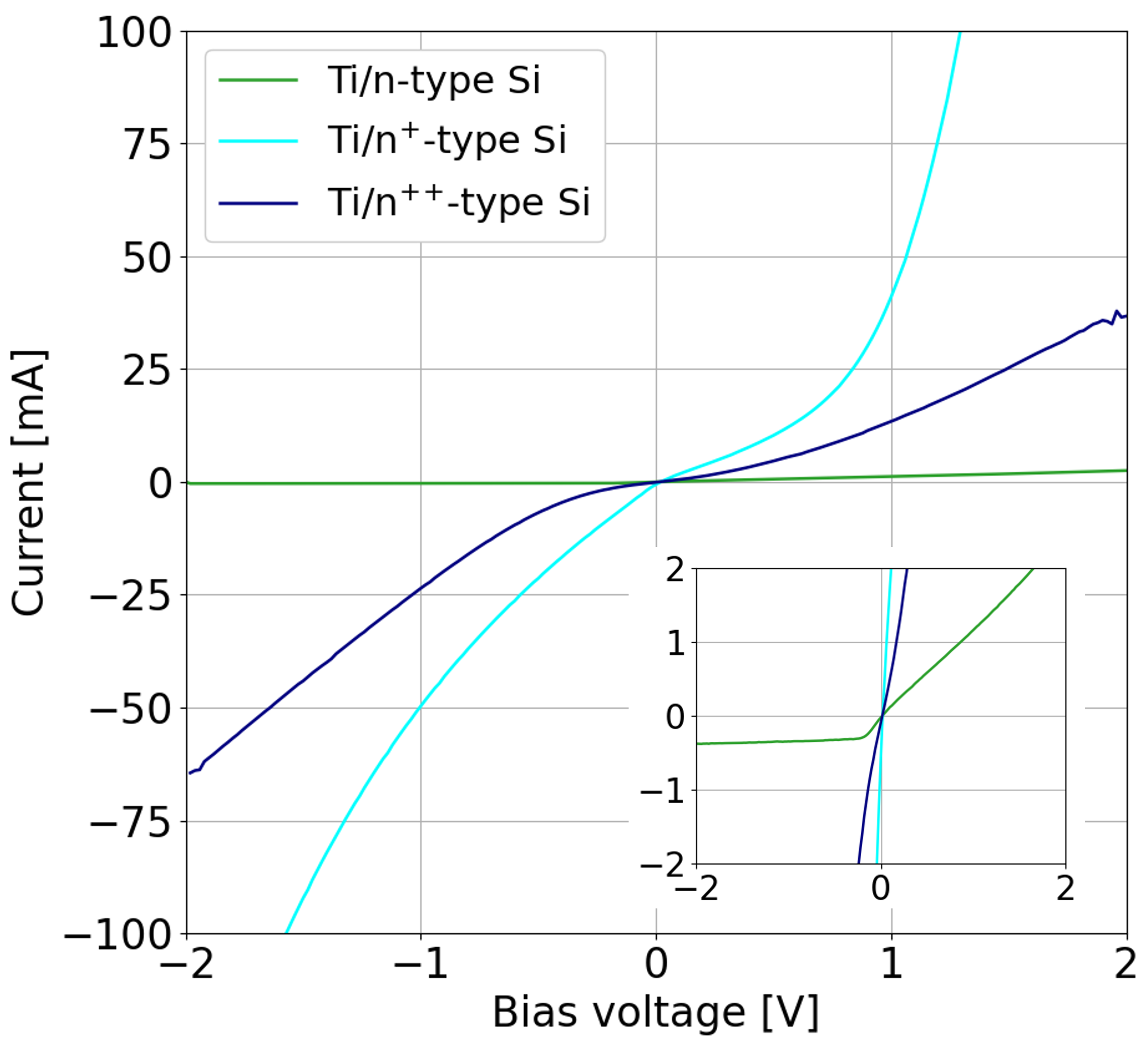}
	   \caption{$I$($V$) for Ti/$n$-doped Si.}
	\end{subfigure}
	\hfill
	\begin{subfigure}{0.235\textwidth}
		\includegraphics[width=\textwidth]{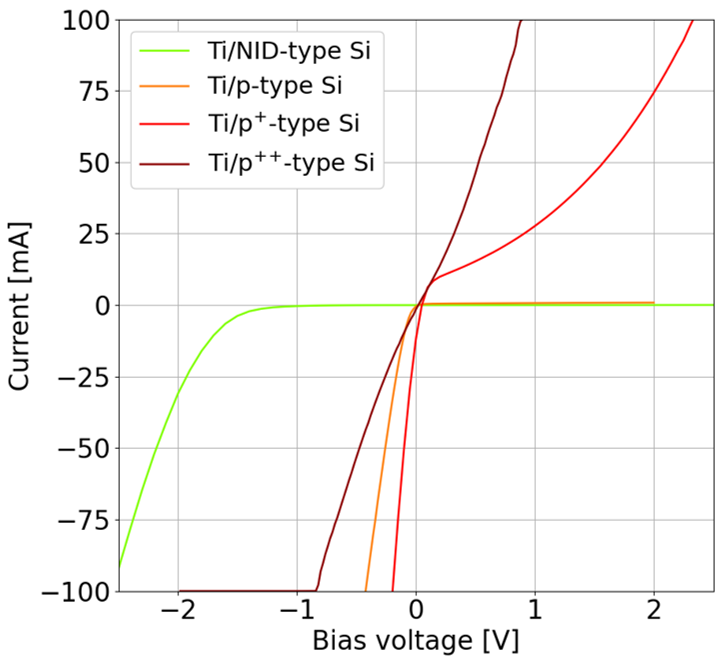}
		\caption{$I$($V$) for Ti/$p$-doped Si.}
	\end{subfigure}
	\caption{Measured $I$($V$) caracteristics for Ti/$n$ and $p$-doped samples.}\label{I_V_N_P}
\end{figure}

The I(V) characteristics measured for all the samples are shown in Figure \ref{I_V_N_P}, where it can be clearly seen that no rectifying effect is present for $n^+$/$p^+$ and $n^{++}$/$p^{++}$ doping levels, contacts tend to become ohmic.

	\subsection{$I$($V$, $T$) measurements}\label{I_V_T_subsection}

The Schottky barrier height $\Phi_B$ can be derived from the $I$($V$,~$T$) curves. Indeed, the saturation current $I_s$ can be expressed as a function of the barrier height $\Phi_B$ and temperature $T$ of the Schottky barrier \citep{werner_temperature_1993} and the logarithmic $I_s$ value is directly proportional to the quantity $q(\Phi_B - \Delta \Phi)$ where $\Delta \Phi$ is the barrier lowering in eV:

\begin{equation}
I_{\text{s}} = S\,A^{*}\,T^2\,e^{-\frac{q (\Phi_B - \Delta \Phi) }{nkT}}
\label{Saturation_curr}
\end{equation}

with $S$ the area of the device in [$\text{m}^{2}$], $A^{*}$ the Richardson constant in [$\mathrm{A\cdot \text{cm}^{-2}\cdot K^{-2}}$] and $n$ the ideality factor.   

In the case of temperatures when $T$~>~150~K, the ideality factor tends to $n=1$, and the barrier lowering to $\Delta \Phi = 0$. This leads to:

\begin{equation}
\log{\left( \frac{I_{\text{s}}}{S\,A^{*}\,T^2} \right)} = -\frac{q\,\Phi_B}{kT}
\label{I(V,T)_resu}
\end{equation}

\begin{figure}[h]
		\centering
		\includegraphics[width=0.48\textwidth]{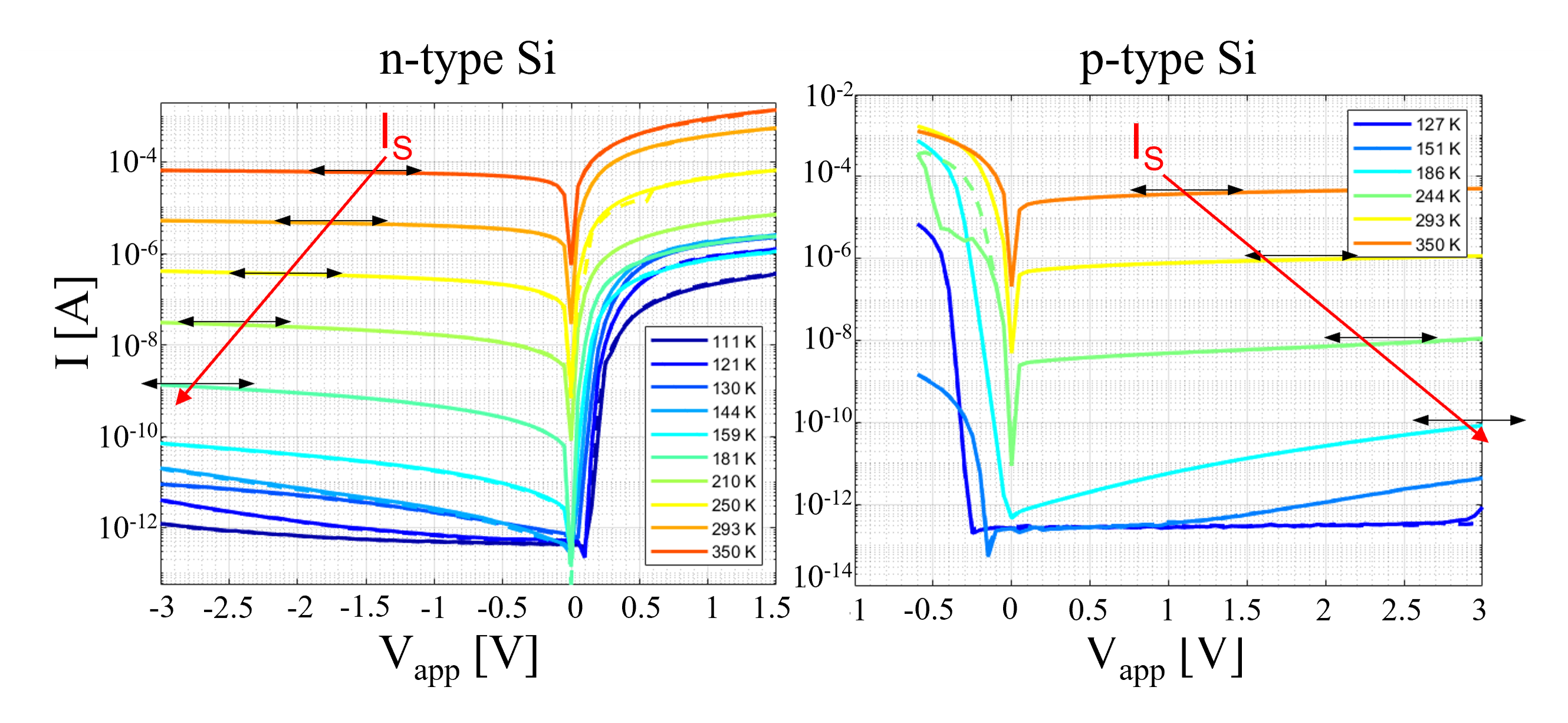}
		\caption{$I$($V$) characteristics measured for different temperatures for lightly doped samples.}\label{I(VT)}
\end{figure}

This equation can then be used with the $I$($V$) measured at different temperatures to obtain the value $\Phi_B$. Figure \ref{I(VT)} shows these curves for temperatures ranging from 111 to 350 K.

The values obtained for $\Phi_B$ are provided in Table \ref{eMeas_table} only for ($n$, $p$) light doping levels. For high doping levels, the space charge width $W$ is too small and the saturation current vanishes.

		\subsubsection{Doping level, space charge width and barrier height of Ti/doped Si interfaces}\label{sec:CV_results}
		
The $C$($V$) measurements allow one to determine the density of dopants $n_D$, the built-in potential $V_B$ and to the space charge width $W$. The methodology consists of measuring the electric capacitance \textit{C} at the interface induced by the space charge area as a function of the bias applied $V$ \citep{werner_temperature_1993, ray_defect_2014}. Indeed, Poisson's equation, in the approximation of the planar capacitor, gives $W$ and $C$ versus $V$, $n_D$ and the built-in potential $V_B$ as expressed by:

\begin{equation}
W(V) = \sqrt{ \frac{2 \varepsilon\, \left( V_B - V \right)} {q n_D } }
\label{W(V)}
\end{equation} 

\begin{equation}
C(V) = \frac{\varepsilon\ S}{W} = S \sqrt{ \frac{\varepsilon\ q n_D}{2 \left( V_B - V \right)} }
%\frac{1}{C^2} = \left( \frac{2}{S^2\,\varepsilon\, q} \right) \frac{1}{n_D} \left( V_B - V \right)
\label{C(V)}
\end{equation} 

\begin{figure}[h]
	\centering
	\includegraphics[width=0.48\textwidth]{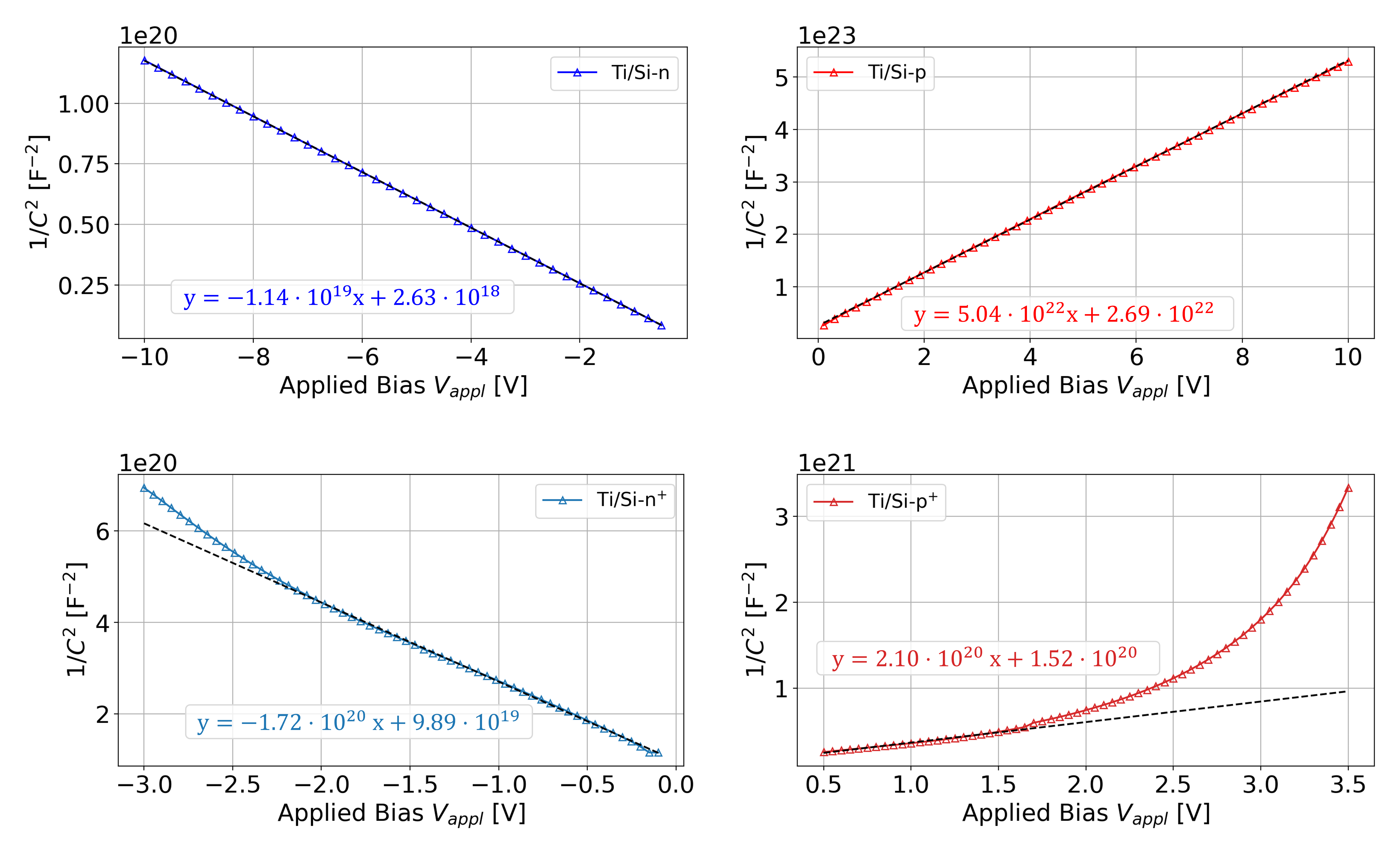}
	\caption{$C$($V$) measurements for: lightly $n$-doped and $p$-doped Si (top-left and top right), and for intermediate doping level (bottom).}\label{C(V)_1}
\end{figure}

Figure \ref{C(V)_1} provides the $1/C^2$ curves as a function of the applied bias for measurements at a frequency of 100 kHz. For ($n$, $p$) doping levels, curves are linear as predicted by Equation \eqref{C(V)}. For ($n^+$, $p^+$) doping levels, only small bias give linear comportement of $1/C^2$. The highest doping levels ($n^{++}$, $p^{++}$) do not present space charge width and no capacity measurement could be performed on these samples.

\begin{figure}[h]
	\centering
	\includegraphics[width=0.48\textwidth]{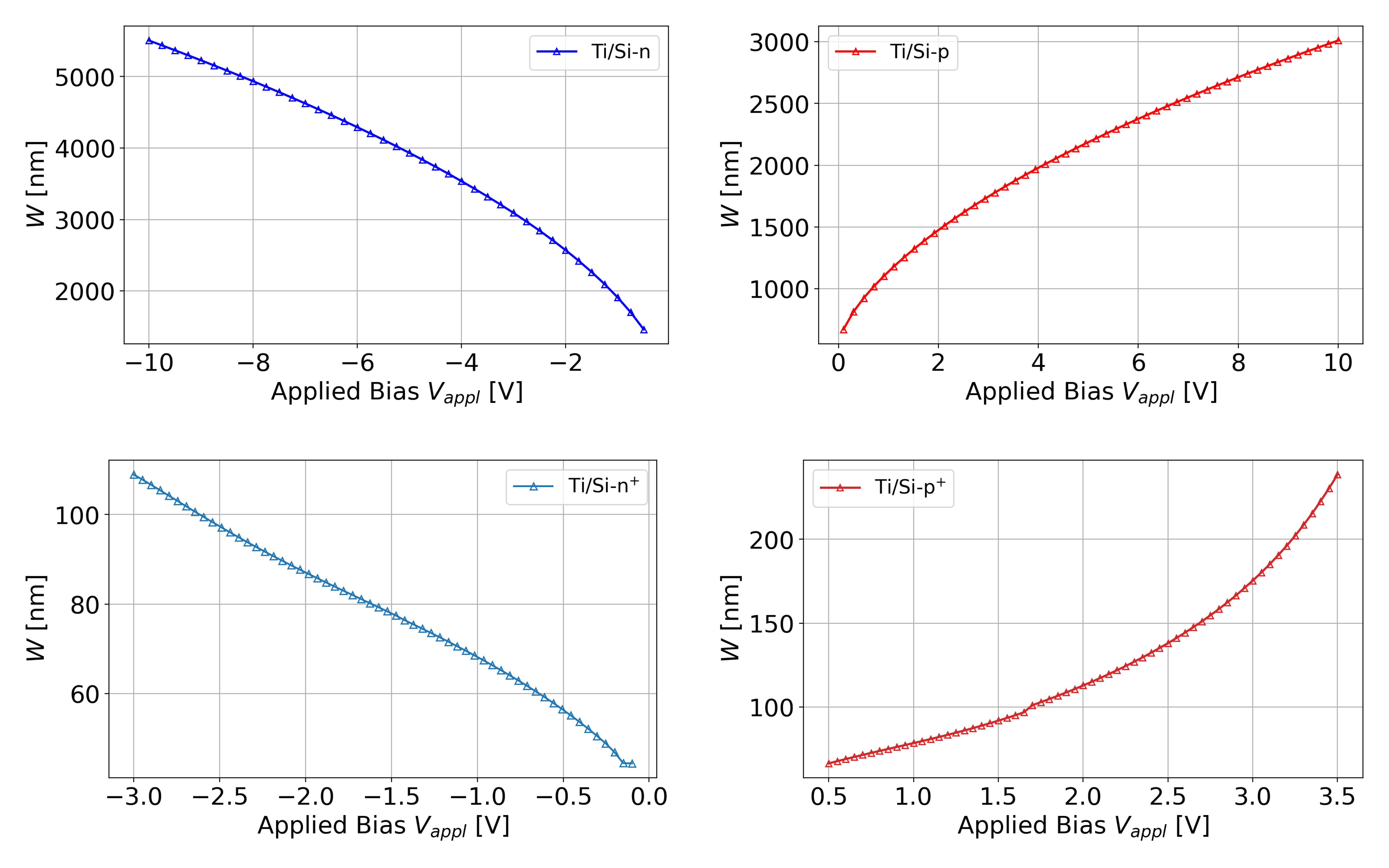}
	\caption{Space charge width $W$ as function of the applied bias obtained from $C(V)$ measurements presented in Figure \ref{C(V)_1}.}\label{W_V}
\end{figure}

From Equation \eqref{C(V)}, the sloped and intercept of the curve $1/C^2$ have been derived to determine the dopant concentrations $n_D$ and the values of the built-in potential $V_B$. The Schottky barrier height $\Phi_B$ is then obtained form $\Phi_B~=~(E_C~-~E_f)~+~qV_B $.
Figure \ref{W_V} provides the resulting variation of the space charge width $W$ versus the applied bias obtained from Equation \eqref{W(V)}. These values are in good accordance with those simulated for a 2V bias voltage (refer to Section \ref{Sample_design_sec}).

The results obtained from the two $I(V,T)$ and $C(V)$ methods are summarized in the Table \ref{eMeas_table}. 

\begin{table}[h]
	\caption{Table of doping level, space charge width and barrier height estimated by the electrical measurements.%\newline
 %   \footnotesize{$^{\text{[a]}}$: value taken based on electrical conductivity provided by fabricant.}\newline
 %	   \footnotesize{$^{\text{[b]}}$: N/D are for Not Determined values.}
    }\label{eMeas_table}
\centering
\small
%	\begin{tabular}{ p{0.08\textwidth}<{\centering}  p{0.10\textwidth}<{\centering}  p{0.08\textwidth}<{\centering} p{0.08\textwidth}<{\centering} p{0.08\textwidth}<{\centering}}	
	  \begin{tabular}{|c|c|c|c|c|}	

		\toprule
        $\,$ Doping type $\,$ &    \makecell{$\,$ $n_{\text{D}}\,[\text{cm}^{-3}]$ $\,$\\ $C(V)$} &  \makecell{$\,$ $W\,[\text{nm}]$ $\,$ \\ $(@\pm2 V)$}  &    \makecell{$\,$ $\Phi_{B}\,[\text{eV}]$ $\,$\\ $I(V,T)$} &   \makecell{$\,$ $\Phi_{B}\,[\text{eV}]$ $\,$\\ $C$($V$)} \\
        		\hline
		$n$-type			&	$4.4\cdot 10^{14}$	&	$2275$ &    $0.31$  & $0.54$\\
		$p$-type			&	$1.5\cdot 10^{15}$	&	$1450$ &    $0.45$  & $0.75$\\
		$n^{+}$-type		&	$4.4\cdot 10^{17}$	&	 $85$ &   - & $0.559$\\
		$p^{+}$-type		&	$4.0\cdot 10^{17}$	&	$110$ &   - & $0.417$\\

		\toprule
	\end{tabular}
\end{table}

The values of $n_D$ are in the range of values provided by the wafer producer, and the values of $W$ correspond to those given by the COMSOL simulations and given in Section \ref{Sample_design_sec}. \\

Concerning $\Phi_{B}$ estimations, both techniques give different values. Indeed, at low temperature, the $I(V,T)$ are obviously impacted by generation-recombinaison leakage current, and SBH extraction are thus less accurate. $C(V)$ measurements were performed in frequencies ranging from 100 Hz to 2 MHz and extended voltage range. Doping densities and barrier heights were extracted from the 1/C2 plots in a frequency range were no significant variation was observed, in order to avoid influence of interface traps. The $\Phi_{B}$ values are consistent with those of literature, i.e. $\Phi_{B}$ = 0,5 eV for Ti/$n$-Si and $\Phi_{B}$ = 0,61 eV for Ti/$p$-Si as given in Sze \citep{sze2007physics}.

 \section{Thermal Boundary Conductance measurements}\label{TBC_meas_sec}

Two metal/semiconductor interfaces were thermally studied: Ti/Si and Pt/Si. The TBC was measured on both samples sets but only the Ti/Si interfaces were investigated with polarization. Before TBC measurements, the thermal properties of silicon was studied to improve the estimation of TBC.

    \subsection{TBC measurements}\label{TBC_measurement}

The TBCs at Ti/Si interfaces were measured using the frequency domain photothermal radiometry (FD-PTR \citep{horny_kapitza_2016, hamaoui_electronic_2018, hamaoui_spatially_2020}). The principle of FD-PTR relies on sample laser heating and measurement of the surface thermal radiation to determine its thermal properties. In the setup used, the laser ($P_{las}~=~ 500$\,mW, $\lambda~=~457$\,nm, model: PhoXx-457-500) is modulated in amplitude at frequency $f$, and focused on a diameter spot size 96.3 µm ($1/e^2$). The infrared emission of the surface of the sample is collected by an HgCdTe detector ($\lambda_{max}$ = 11 µm, $\lambda_{range}$ = 5-14 µm, model: Kolmar technologies HMPV 11-1-J1/DC Ge). A Zurich Instrument ZI-UHFLI600 lock-in amplifier (LIA) is employed to measure the radiometric signal $RS$, which is taken, in the first approximation to be proportional to the complex alternative surface temperature $T_{AC}^{*}$: 

\begin{equation}\label{EqRS}
RS \propto T_{AC}^{*} = T_{AC} \cdot e^{i \varphi_{AC}}
\end{equation}

Here, the average amplitude of the alternative temperature amplitude $T_{AC}$ is about a few kelvins. A voltage/current source for bias or current application to the sample and monitoring is also used. The estimation of TBC was performed by fitting experimental data with a multilayered heat equation model, including volumetric heat sources, solved by a Gauss-Newton algorithm \citep{horny_kapitza_2016, hamaoui_spatially_2020}. In the minimization process, where only phases were used, all thermal quantities, except the M/SC TBC, were treated as fixed parameters and uncertainties were calculated accordingly to it. 

\begin{figure}
    \centering
    \includegraphics[width=0.45\textwidth]{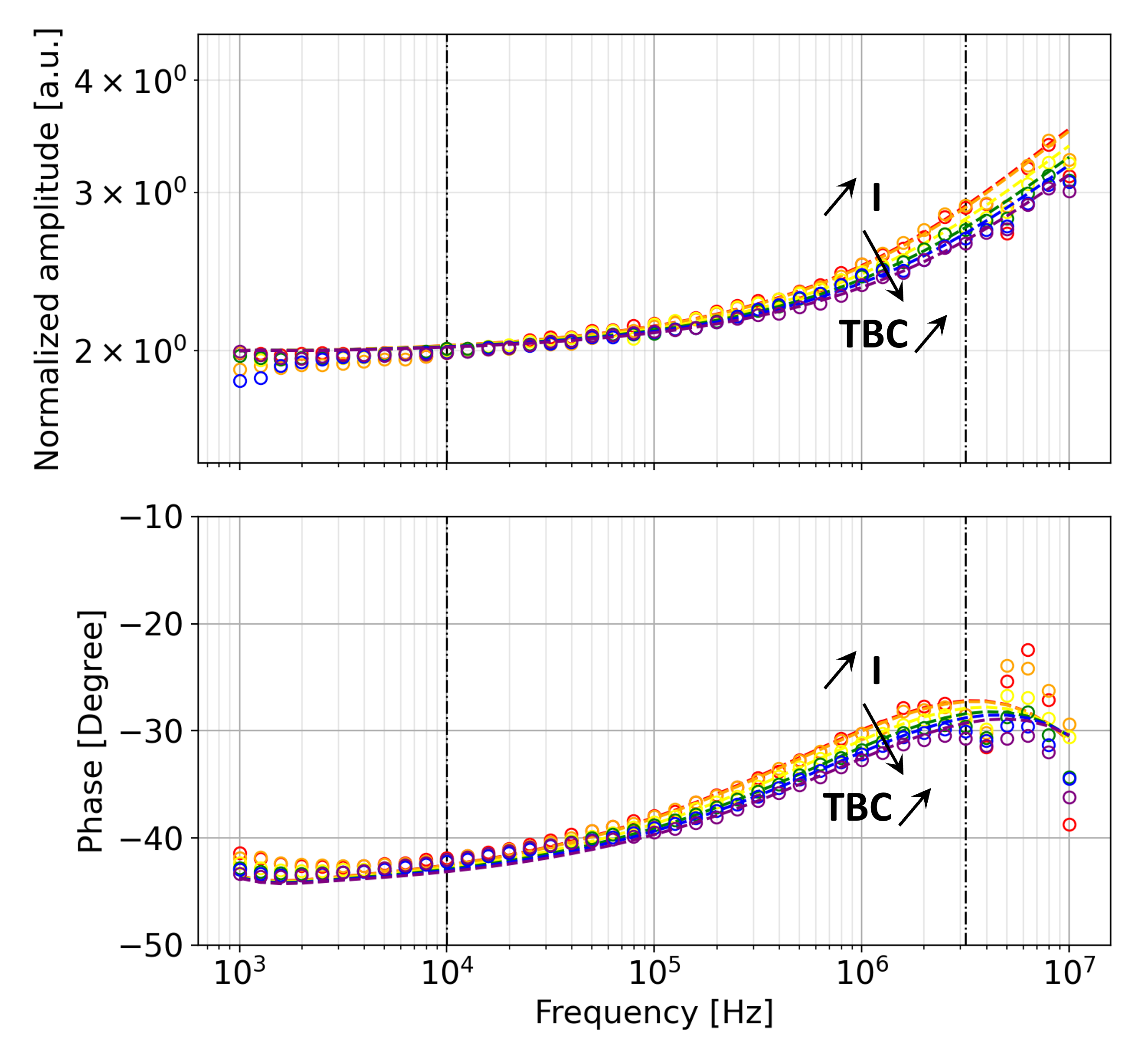}
    \caption{PTR measurements (circles) and model (dashed line) for one Ti/Si sample, showing evidence of the increase of TBC with applied bias.}
    \label{fig:PTR_Result}
\end{figure}

An example of the amplitude and phase of the FD-PTR obtained on Ti/$n$ Si sample is shown in Figure \ref{fig:PTR_Result} where the values of TBC ranges from 62 to 84~$\text{MW} \cdot \text{m}^{-2} \cdot \text{K}^{-1}$. The vertical lines indicate the frequency range over which the parameter estimation was performed.

\begin{figure}
    \centering
    \includegraphics[width=0.45\textwidth]{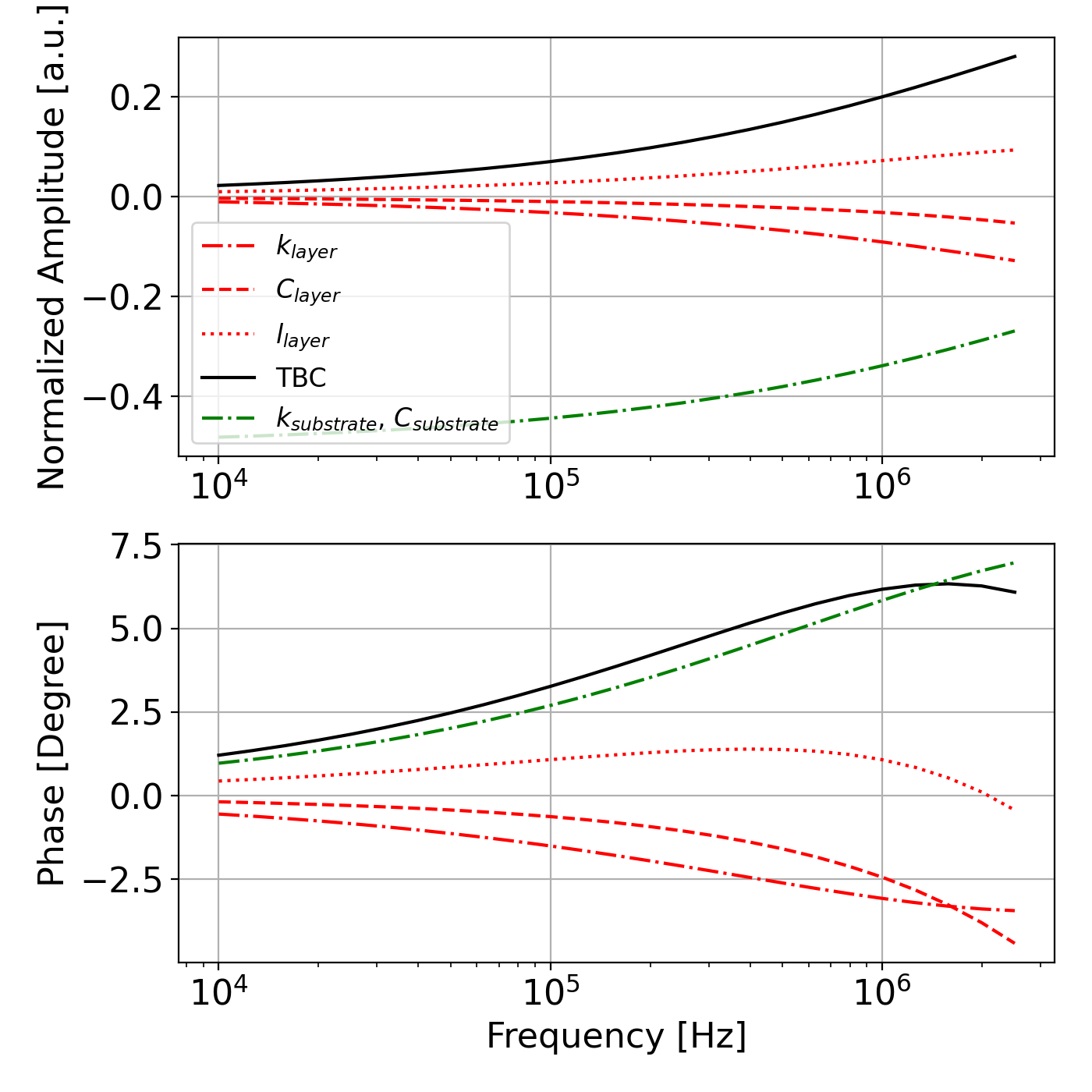}
    \caption{Sensitivities of the thermal model to the different parameters, including TBC.}\label{fig:sensitivities}
\end{figure}

The observed variations of phases are around $5$ degrees close to the maximum of sensitivity in a frequency range near 1 MHz, as shown in the sensitivity curves on Figure \ref{fig:sensitivities} . 

It is also clear that the sensitivity to the thermal conductivity of the silicon substrate is of the order of that of TBC and could imply significant errors, resulting in a major deviation of identified TBC values. For the studied samples and the configurations of measurements, the thermal conductivity of silicon depends on the doping level and on temperature. These dependences were included in the inverse problem to increase the precision of the TBC estimations.
The variations of the thermal conductivity of the silicon substrate with doping level and temperature are addressed in the next two sections.

	\subsection{Silicon thermal conductivity}
	
    \subsubsection{Doping level dependency}

\begin{figure}[h]
	\centering
	\includegraphics[width=0.45\textwidth]{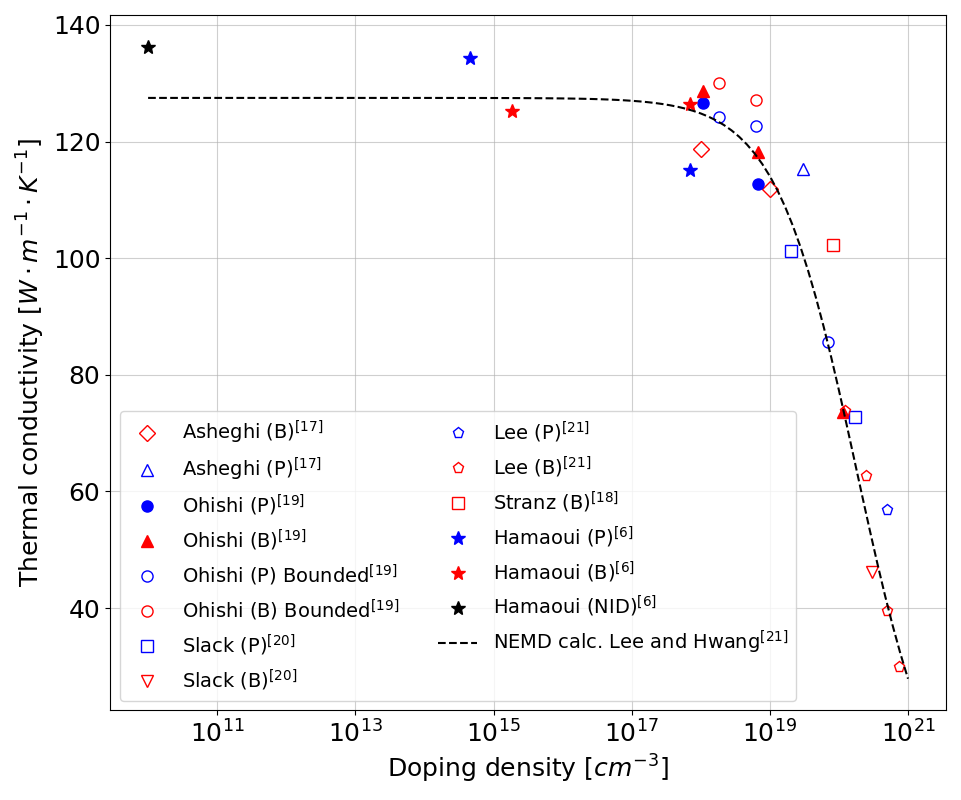}
	\caption{Thermal conductivity of doped silicon as function of substrate dopant concentration taken from different literature works\cite{asheghi_thermal_2002, stranz_thermoelectric_2013, ohishi_thermoelectric_2015, hamaoui_electronic_2018}. Along with these measurements, the corrected-NEMD made by Lee and Hwang\citep{lee_mechanism_2012} is plotted.}\label{fig:kSi_vs_Dop_3omega}
\end{figure}

In our TBR identification process, the values of the doped silicon thermal conductivity $\kappa_{\text{Si}}$ at room temperature as a function of doping level were derived from previous measurements [Asheghi \textit{et al.}\citep{asheghi_thermal_2002}, Stranz \textit{et al.}\citep{stranz_thermoelectric_2013}, Ohishi \textit{et al.}\citep{ohishi_thermoelectric_2015}, Slack \textit{et al.}\citep{slack_thermal_1964} and Hamaoui \textit{et al.}\citep{hamaoui_electronic_2018}] and atomistic simulations [Lee and Hwang \citep{lee_mechanism_2012}].
As shown in Figure \ref{fig:kSi_vs_Dop_3omega} reference data were fitted according to \citep{lee_mechanism_2012}:

\begin{equation}
	\kappa = \frac{\kappa_{\text{Si}}}{1+ A \left( \frac{n_{D}}{10^{20}}\right)^{\alpha}}
	\label{Lee_eq}
\end{equation}

where $A$ and $\alpha$ are fitting parameters.

$\kappa_{\text{Si}}$ was set to the average value corresponding to measurements of the thermal conductivity of intrinsic and light doped silicon substrates $\kappa_{\text{Si}}~=~126~\,\text{ W}\cdot~\text{m}^{-1}\cdot~\text{K}^{-1}$ at room temperature.

    \subsubsection{Temperature dependency}

The thermal conductivity of silicon also evolves with temperature as a result of phonon scattering inside the material. In fact, the contributions of heat dissipation inside the material are the following: i) boundary scattering; ii) impurity scattering and iii) anharmonic phonon-phonon scattering. The first occurs at low temperature, giving a small contribution to the thermal conductivity.
At higher temperatures, thermal conductivity becomes dominated by phonon-impurity scattering. Finally, for temperature higher than $300$ K, anharmonic scattering controls the reduction of the thermal conductivity through the decreased phonon mean path.

\begin{figure}
    \centering
    \includegraphics[width=0.4\textwidth]{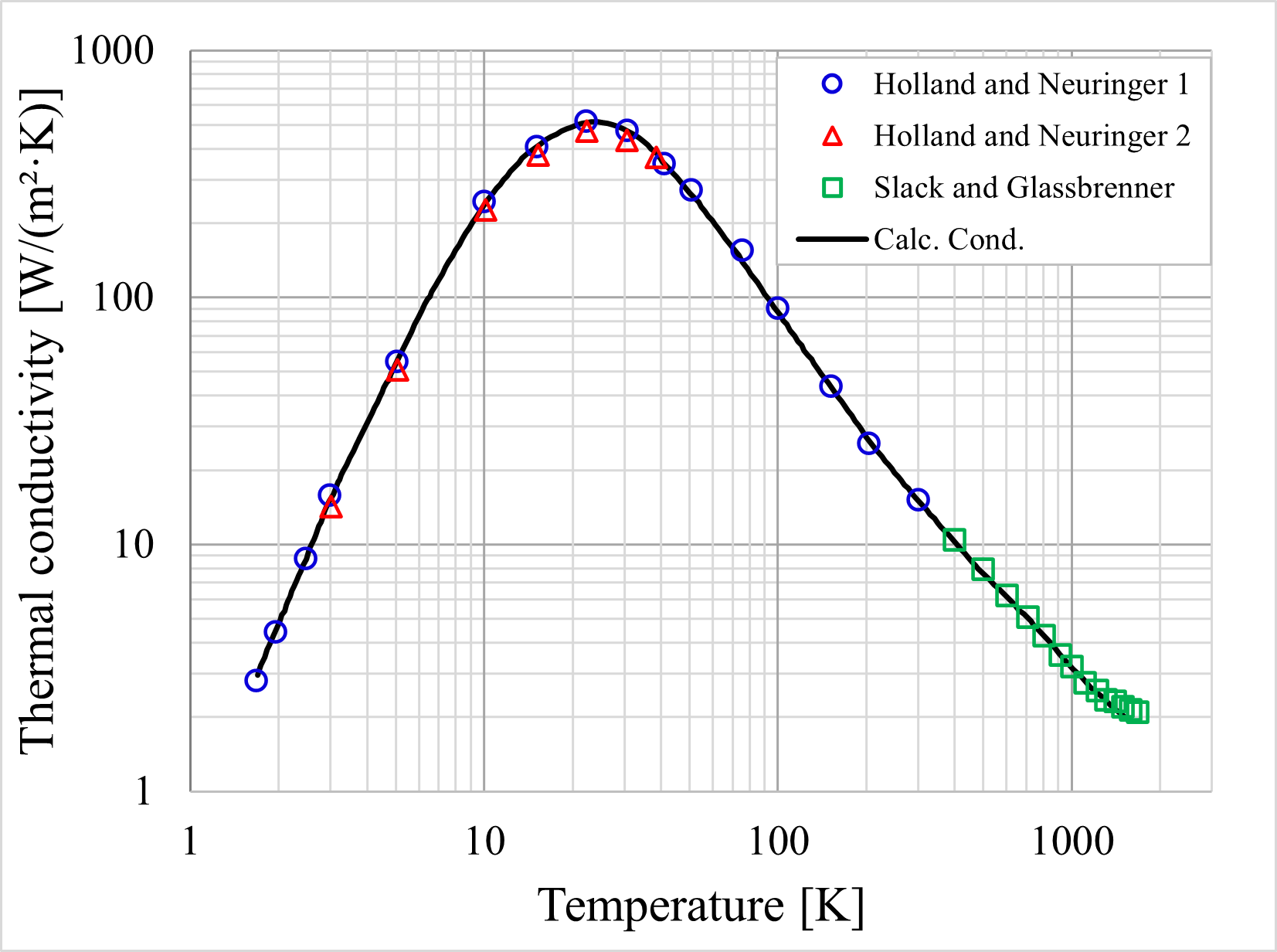}
    \caption{Thermal conductivity of silicon evolution with the temperature, taken from Holland and Neuringer\citep{holland_analysis_1963} and Slack and Grassbrenner\citep{glassbrenner_thermal_1964}.}
    \label{fig:kSi_vs_T}
\end{figure}

Figure \ref{fig:kSi_vs_T} presents the relative contribution of these channels in the temperature range of 1 to 1000 K taken from the literature\citep{holland_analysis_1963, glassbrenner_thermal_1964}. 

The doping level dependence as well as the temperature dependence of doped silicon thermal conductivity $\kappa_{\text{Si}}$ were included in the inverse problem to increase the accuracy in the estimation of the TBC.

    \subsection{TBC as function of the Si doping level – no polarization bias}\label{TBC_vs_dop_section}
	
The doping level of the substrate could possibly affect the TBC at M/SC interface. The doping process produces defects in the crystal, and the presence of these defects can potentially alter the carrier transfer behavior at the interface and affect channels 1 and 2 of heat transmission (Figure \ref{fig:pathways}) by modifying the vibrational density of state (vDOS), a part of the $G$ calculation according to theoretical work \cite{gordiz_phonon_2016, feng_unexpected_2019, lindsay_perspective_2019, giri_review_2020}. It could also modify channel 3 via the modification of the electronic carrier density with the variation of the space charge width $W$.

Mesurements of TBC were performed on samples of $100$ nm Ti and Pt metal films on Si substrates with the different doping levels given in Table \ref{data_table}. 

\begin{figure}[h!]
	\centering
	\includegraphics[width=0.45\textwidth]{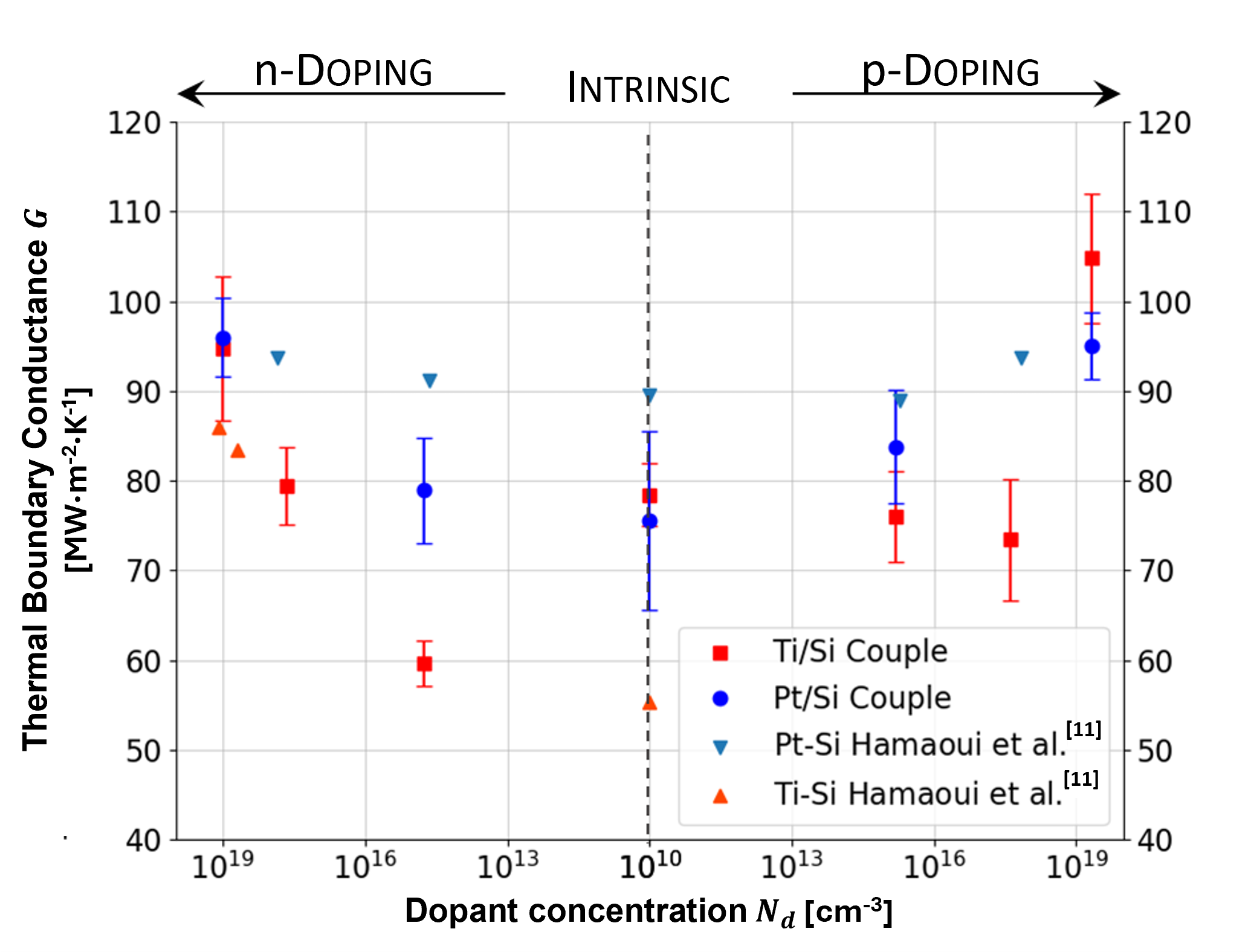}
	\caption{Thermal boundary conductance as a function of the substrate doping concentration.}\label{fig:GvsDop} 
\end{figure}

Figure (\ref{fig:GvsDop}) reports on the thermal boundary conductance measured as a function of the doping level of the two M/SC couples. We have also compared these measurements to previous data published in the literature.

For the Ti/Si system, we can observe a slight decrease of the TBC with the concentration in the regime of light to moderate doping. By contrast, at higher doping levels, the TBC increases. Two mechanisms may explain this increase: first, the collapse of the space charge region; secondly, the increase of the charge density carrier within the SC.

	\subsection{TBC as function of the electrical bias}
	
We have also explored the influence of a voltage bias or an electrical current on the TBC, depending on the polarization direction, i.e. direct vs inverse. The aim was to clearly evidence the role of thermally excited charges in thermal transport at metal/semiconductor interfaces. 
As Schottky diodes present a direct and a rectifying mode, we have represented the results in two ways: the first is a TBC versus applied voltage plot in reverse mode, and the second is TBC versus electrical current in the direct mode. In the case of lightly doped semiconductors ($n$- and $p$-type), both depicted representations are possible, due to the weak electrical current in reverse mode of the junction. However, because the highest doping levels present a higher leakage current due to a smaller space charge width $W$, a current representation was used, even for the reverse mode.

\begin{figure}[h]
		\centering
		\includegraphics[width=0.45\textwidth]{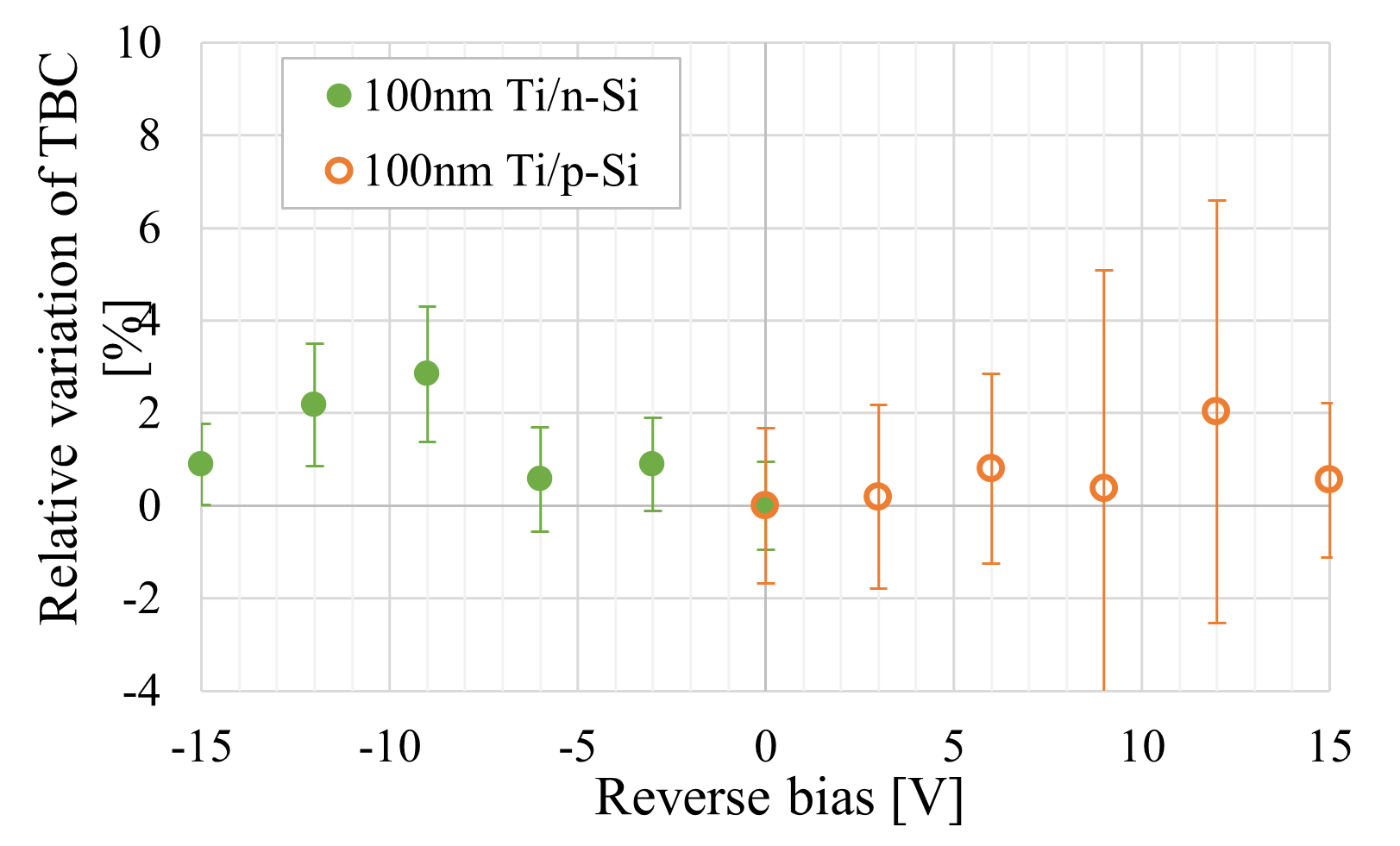}
		\caption{TBC as function of the applied reverse bias for $n$- and $p$-doped Si ($n_D = 0.6-1.6\cdot 10^{15}\,\text{cm}^{-3}$).}\label{TBC_reverse_1}
\end{figure}

\begin{figure}
	\centering
	\includegraphics[width=0.45\textwidth]{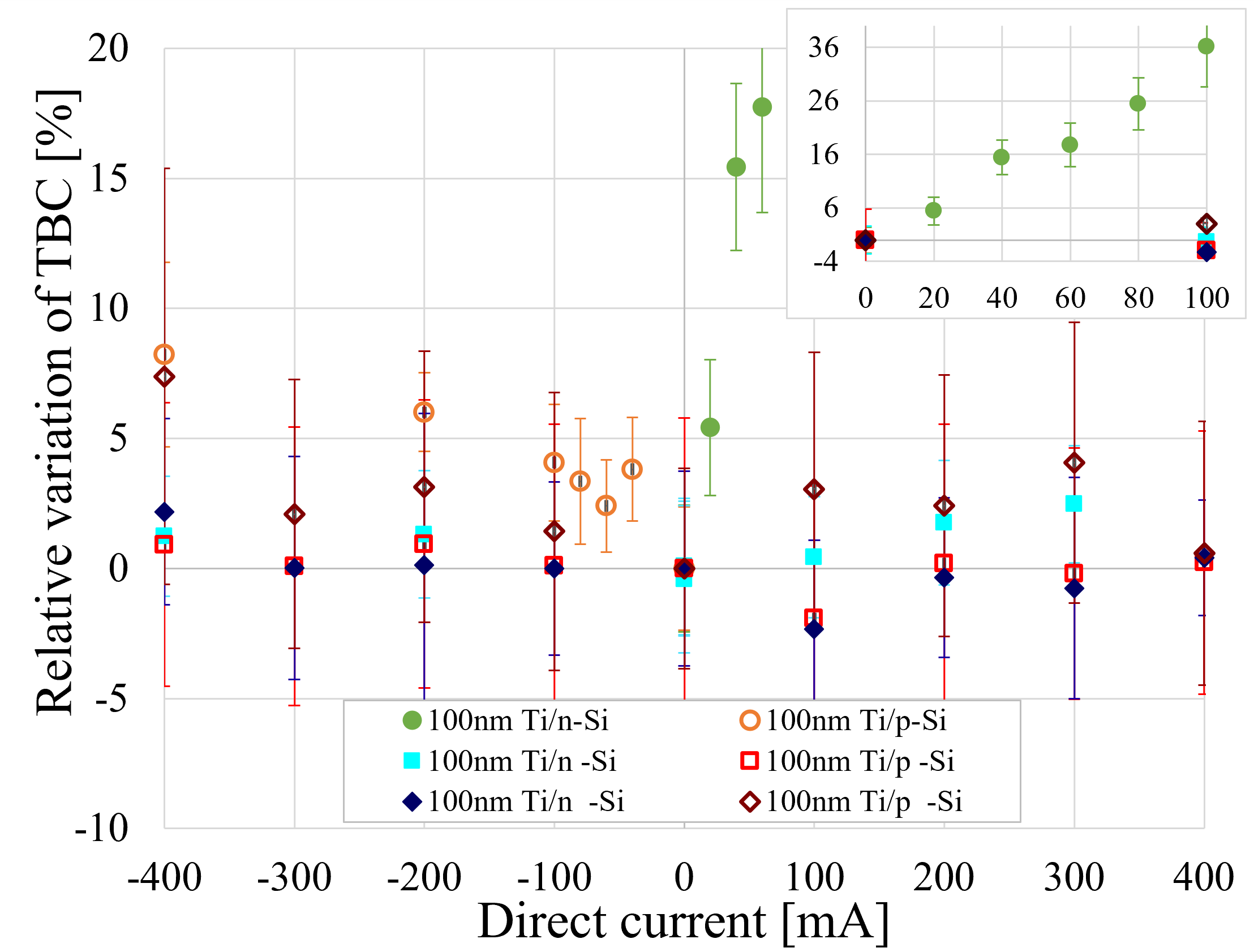}
	\caption{TBC as a function of current, in the diode regime in direct and high leakage reverse current junctions. The inset represents a zoom of the positive current region displaying the full scale.}\label{Final_figure_TBC_vs_Curr_relative}
\end{figure}

The only samples that showed a clear rectifying effect were the $n$- and $p$-type (corresponding to light doping levels). The relative variations of the TBC is represented as a function of the applied reverse bias in Figure \ref{TBC_reverse_1}. It is important to note that on the figures \ref{TBC_reverse_1} and \ref{Final_figure_TBC_vs_Curr_relative}, the contributions of parameters of the model assumed to be known are the same, except the variations with temperature which are accounted. The uncertainties shown on the relative variations of TBC are thus only the uncertainties due to the noise.

As seen in Figure \ref{W_V}, the space charge width $W$ increases with the applied bias (approximately 300 nm/V and 500 nm/V for $p$ and $n$, respectively), so it is clear that the increase in $W$ does not affect TBC more than 3$\%$ at light doping levels. It is also important to note that the variations are always positive.

For these two samples, the variation of TBC for direct mode can also be represented versus current. For all of the other samples, the TBC values are given also as a function of the direct current as well as of the high leakage reverse current. These results are shown in Figure \ref{Final_figure_TBC_vs_Curr_relative}. The value of the electrical current is limited to $I$ = 500 mA to prevent any damage to devices even if the probe area is large compared to what is found in the SC industry ($j \approx \mathrm{2.5 A/cm}^{2}$ compared to the $j >$ 10 A/{cm}$^{2}$ of µ-LED chips\citep{chen_mechanism_2021}). Here, the value of the current density is moderate and below the typical value leading to material's damage. However, as observed for any electrical device, the current induced in the sample generates Joule heating. The increase in temperature was measured with a K-type thermocouple during the PTR experiment for absolute DC temperature measurements, and the variation of the substrate thermal conductivity was also accounted for. The highest levels of heating amounts to 50 K for light doping levels. The corresponding evolution of the TBC with the current density is shown in Figure \ref{Final_figure_TBC_vs_Curr_relative}.
    
The observed increase in TBC is more pronounced for lightly doped systems, specifically for the $n$-type with an increase close to $40 \%$. However, this increase corresponds to a low TBC (corresponding to point at $G = 60\,\text{ MW}\cdot\text{m}^{-2}\cdot \text{K}^{-1}$ in Figure \ref{fig:GvsDop}). This could be explained by the collapse of the space charge region and the subsequent increase of the electron-SC charge coupling factor $G_{\text{e}^{M}-\text{e/h}^{SC}}$. Other samples exhibit an increase of the TBC close to $8\%$. The TBCs of the other samples turn out to be constant and weakly dependent on the direct current, although always positive.

\begin{figure}[h]
	\centering
	\begin{subfigure}{0.48\textwidth}
		\centering
		\includegraphics[width=0.65\textwidth]{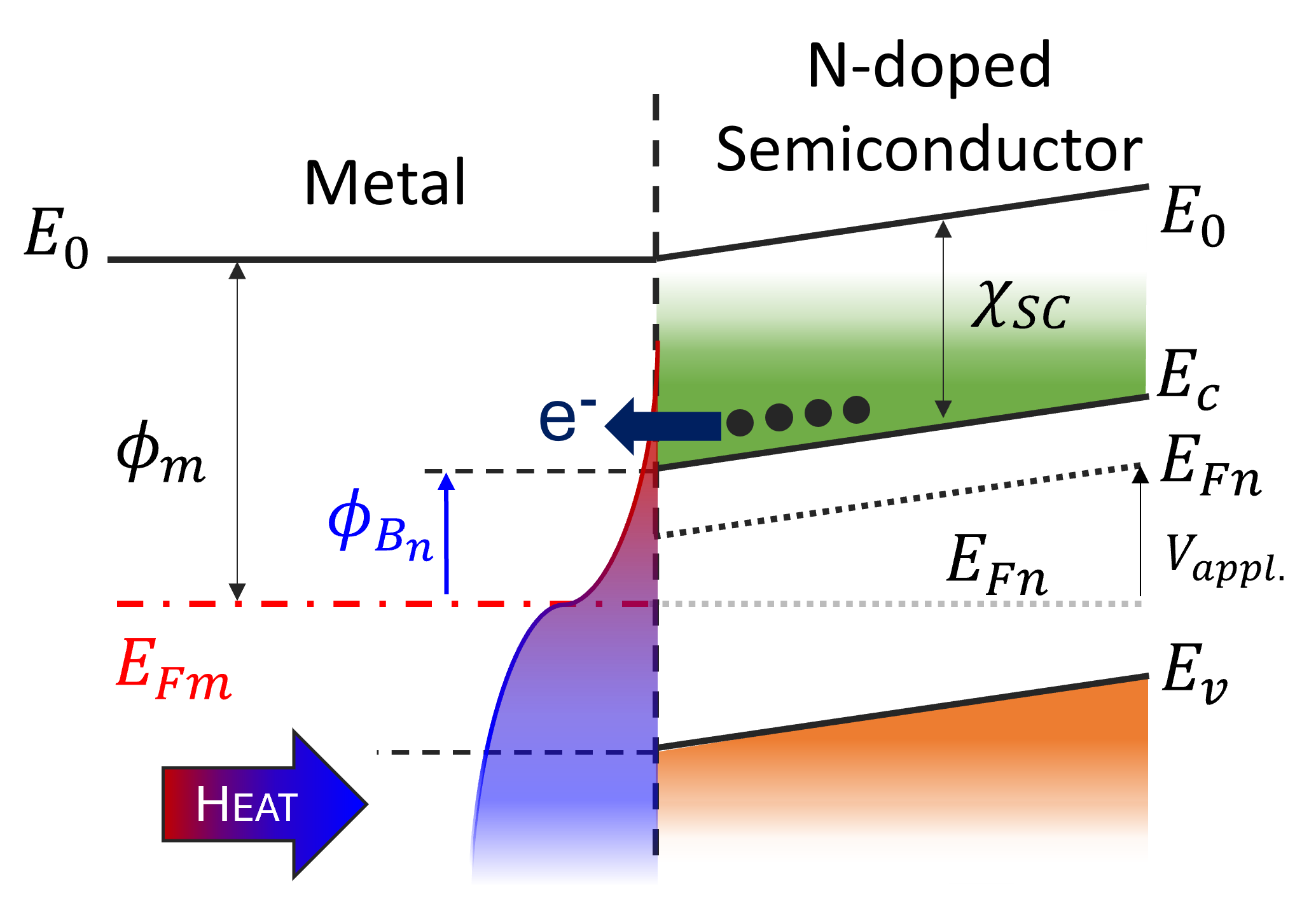}
		\caption{Band diagram for Ti/$n$-doped substrate in the direct mode (corresponding to a positive polarization).}
	\end{subfigure}
	\hfill
	\begin{subfigure}{0.48\textwidth}
		\centering
		\includegraphics[width=0.65\textwidth]{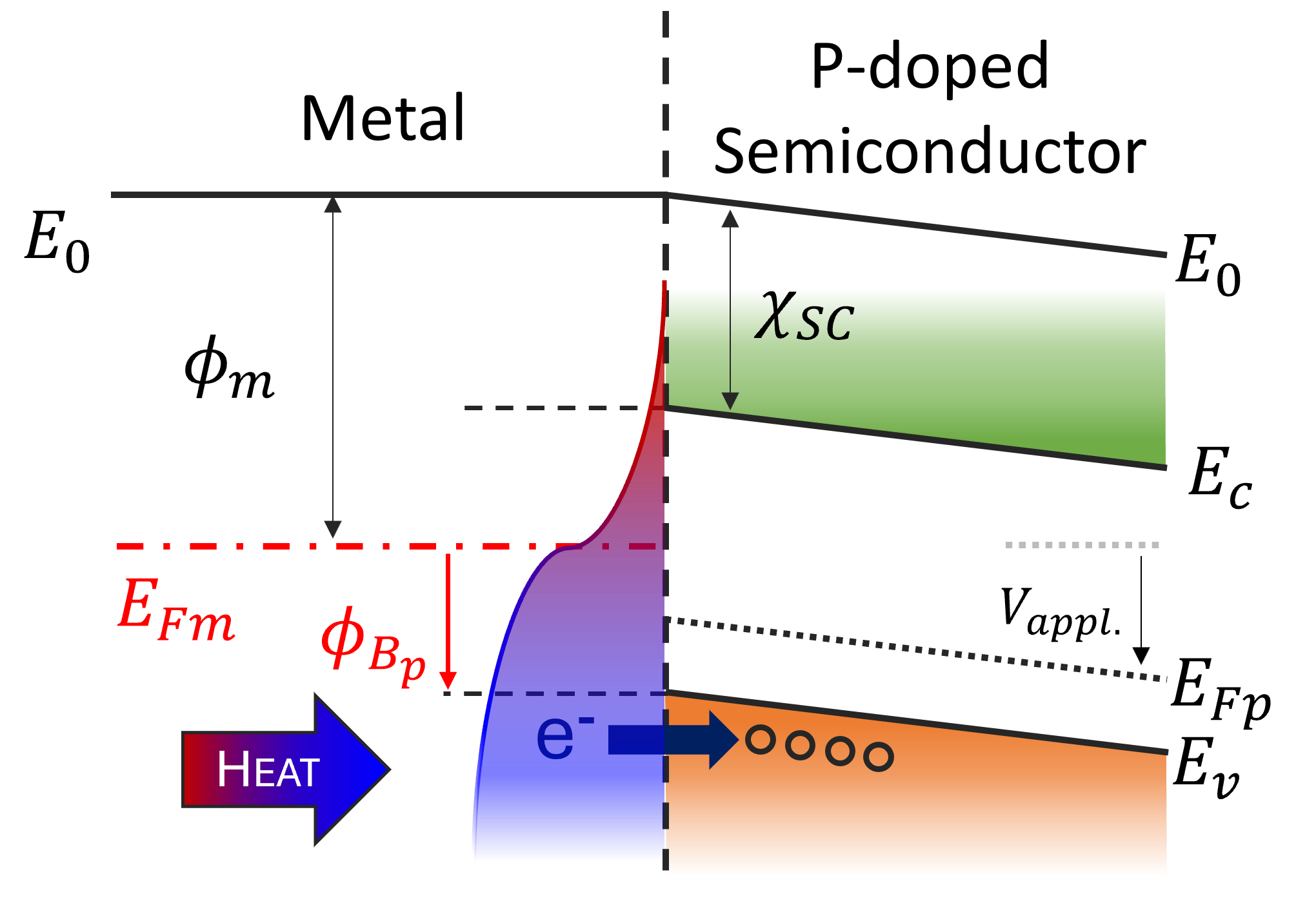}
		\caption{Band diagram for Ti/$p$-type substrate in the direct mode (negative polarization).}
	\end{subfigure}
	\caption{Illustration of the charge flow across the junction in the regime of direct bias M/SC diode.}\label{Band_diag_direct}
\end{figure}

To illustrate the influence of conduction electrons on heat transfer at interfaces, it is interesting to represent the different band diagrams, as shown in Figure \ref{Band_diag_direct} for direct-mode Schottky diodes with electrical charge flux. These two configurations correspond to the highest increase of TBC.

In the FD-PTR configuration, as in all pump/probe experimental setups, the heat source acts on the metallic film, the transducer, and thus heat flows from the metal side to the semiconductor. However, the main increase of TBC, i.e. for Ti/$n$ Si, occurs for charges travelling from the SC to the metal. Therefore, the free electrons can not contribute to the interfacial heat flux and do not contribute to the TBC. For the two other enhancements of TBC, i.e. for Ti/$p$-Si and Ti/$p^{++}$-Si, electrons that contribute to the current are cold electrons and thus do not carry heat. Consequently, the main contribution to the increase of TBC is the reduction of the space charge area $W$ due to the increase of dopant concentration.

\section*{Conclusion}

In this study, the thermal boundary conductance (TBC) at metal/doped semiconductor interfaces under different electrical conditions was investigated by using photothermal radiometry. Of particular interest here is the effect of an electric current applied across the titanium/silicon junction. To determine the  value of the intrinsic TBC, the measurements were corrected for the increase of the sample DC temperature due to Joule heating. In addition, the variations of thermal conductivity of the doped silicon substrates with the doping level was taken into account. With these corrections, an enhancement of 20 \% for Ti/Si junctions at high doping levels was reported, whatever the doping type, $n$ or $p$. This enhancement is related to the collapse of the space charge area (SCA) which offers optimal conditions for the charge carriers in the semiconductor to interact with the metal electrons at the junction. Although this decrease of the SCA should be also realized by applying a current, the increase in the TBC is not solely attributable to the reduction of the SCA. At high doping concentrations, however, the SCA has a limited spatial extension, and the TBC saturates when a current is applied across the junction. In the case of ohmic contacts, the slight increase in the TBC reported here can be attributed to the transfer of energy carried by hot metal electrons which thermalize with the lattice after crossing the junction. \\ 
In conclusion, this study highlights the key role of the spatial extension of the space-charge area in the enhancement of the TBC at metal/semiconductor junctions, while charge transfer by hot electrons turns out to play a secondary role.

\section*{Acknowledgement}
This work was carried out thanks to the financial support of the National Research Agency (ANR) PRC program (ANR-20-CE09-0024), and the NANOPHOT graduate school (ANR-18-EURE-0013).
Some of the experiments were carried out within the Nanomat platform (www.nanomat.eu) supported by the Ministère de l'Enseignement Supérieur et de la Recherche, the Région Grand Est, and FEDER funds from the European Community.
This work was supported by the French RENATECH network. This work has been partially undertaken with the support of IEMN fabrication (CMNF) facilities.

%\nocite{*}
\bibliography{2025_11_JAP_Influence_of_electric_properties}% Produces the bibliography via BibTeX.

\end{document}